\documentstyle[12pt]{article}

\def\preprint#1{\gdef\@preprint{#1}}
\def\bce{\begin{center}}
\def\ece{\end{center}}
\def\be{\begin{equation}}
\def\ee{\end{equation}}
\def\bea{\begin{eqnarray}}
\def\eea{\end{eqnarray}}
\def\thint{\int d^3\! x\,\,}
\def\ggtmn{g^{\mu\nu}}
\def\ggbmn{g_{\mu\nu}}
\def\pbm{\partial_\mu}
\def\pbn{\partial_\nu}
\def\pbr{\partial_\rho}
\def\pbs{\partial_\sigma}
\def\tbmn{T_{\mu\nu}}
\def\ggtrs{g^{\rho\sigma}}

\begin{document}
\baselineskip=.285in

\catcode`\@=11
\def\maketitle{\par
 \begingroup
 \def\thefootnote{\fnsymbol{footnote}}
 \def\@makefnmark{\hbox
 to 0pt{$^{\@thefnmark}$\hss}}
 \if@twocolumn
 \twocolumn[\@maketitle]
 \else \newpage
 \global\@topnum\z@ \@maketitle \fi\thispagestyle{empty}\@thanks
 \endgroup
 \setcounter{footnote}{0}
 \let\maketitle\relax
 \let\@maketitle\relax
 \gdef\@thanks{}\gdef\@author{}\gdef\@title{}\let\thanks\relax}
\def\@maketitle{\newpage
 \null
 \hbox to\textwidth{\hfil\hbox{\begin{tabular}{r}\@preprint\end{tabular}}}
 \vskip 2em \begin{center}
 {\Large\bf \@title \par} \vskip 1.5em {\normalsize \lineskip .5em
\begin{tabular}[t]{c}\@author
 \end{tabular}\par}
 \end{center}
 \par
 \vskip 1.5em}
\def\preprint#1{\gdef\@preprint{#1}}
\def\abstract{\if@twocolumn
\section*{Abstract}
\else \normalsize
\begin{center}
{\large\bf Abstract\vspace{-.5em}\vspace{0pt}}
\end{center}
\quotation
\fi}
\def\endabstract{\if@twocolumn\else\endquotation\fi}
\catcode`\@=12

\preprint{DPNU-94-55\\[0.5mm] SNUTP-96-031\\[0.5mm] WU-AP/57/96}

\title{\Large\bf Monopole-Bubble in Early Universe
\protect\\[1mm]\  }
\author{}
\author{\normalsize Yoonbai Kim${}^{(1)}$, Kei-ichi Maeda${}^{(2)}$, 
Nobuyuki Sakai${}^{(2)}$\\[1.8mm]
{\normalsize\it ${ }^{(1)}$Department of Physics, Pusan National University,
Pusan 609-735, Korea}\\[1.8mm]
{\normalsize\it yoonbai$@$top.phys.pusan.ac.kr}\\[1.8mm]
{\normalsize\it ${ }^{(2)}$Department of Physics, Waseda University,
Shinjuku-ku, Tokyo 169, Japan}\\[1.8mm]
{\normalsize\it maeda, sakai$@$cfi.waseda.ac.jp}
}
\maketitle

\renewcommand{\theequation}{\thesection.\arabic{equation}}
\def\gatij{\gamma^{ij}}
\def\gabij{\gamma_{ij}}
\def\ophi{\phi^{a}}
\def\hphi{\hat{\phi}^{a}}

\begin{center}
{\large\bf Abstract}\\[3mm]
\end{center}
\indent\indent
The nucleation and evolution of bubbles are investigated in the model of an
$O(3)$-symmetric scalar field coupled to gravity in the high temperature limit.
It is shown that, in addition to the well-known bubble of which the inside region
is true vacuum, there exists another decay channel at high temperature which is 
described by a new solution  such that a false vacuum region like a global
monopole remains at the center of a bubble. The value of the Euclidean action of
this bubble is higher than that of the ordinary bubble; however, the production
rate of it can be  considerable for a certain range of scalar potentials. 

\vspace{1cm}\noindent
Keyword(s): cosmological phase transition, finite-temperature field theory,
general relativity

\noindent
PACS number(s): 98.80.Cq, 11.10.Wx

\newpage
\baselineskip=15pt

\pagenumbering{arabic}
\thispagestyle{plain}
\setcounter{section}{1}
\begin{center}\section*{\large\bf I. Introduction}\end{center}
\indent\indent
It is widely believed that the present universe was achieved through a number of
phase transitions while the universe has expanded and cooled \cite{KL}. A
particularly interesting possibility is that of a first-order phase transition
which is described by the formation and growth of bubbles \cite{CC}. In the
course of phase transitions in the early universe, two additional ingredients
should be considered: one is the gravitational effect \cite{CD} and the other is
high temperature \cite{Aff,Lin}. The inflationary scenario arose from the
application of these ideas to cosmology is a synthetic attempt to answer the
fundamental cosmological questions \cite{Gut}. These are the study of new phase
bubbles in interiors of an old phase, but the opposite case, old phase remnants
surrounded by the new phase, has also been an interesting subject \cite{MSSK}.
Recently inflation in the core of topological defects was analyzed in
Ref.\cite{Vil,SSTM,Sak}. Except for the last case, the order parameter of the
model has been basically a real scalar field, thereby ignoring the difference
between the continuous symmetry breaking and the discrete one at the time of
bubble nucleation. Only for the evolution and collision of bubbles, this
difference has been taken into proper account. However, the various species of
solitonic defects, e.g., 1. topological or nontopological, 2. global or local,
3. cosmic strings, monopoles or textures, are determined by the continuous
symmetry which the theory of interest holds \cite{Vil2}. Although soliton
spectra involve complexities, the scenario of bubble nucleation is a simple one
based on one real scalar order parameter, and continuous symmetry does not play
a role for the formation of bubbles \cite{Col}, whether it is due to quantum
tunneling at zero temperature \cite{CC} or due to thermal fluctuation at high
temperature \cite{Aff,Lin}. Therefore it may be interesting to ask whether the
structure of bubbles is affected by internal symmetry from the time of
nucleation or not, and particularly how the internal symmetry coagulates a
matter droplet inside the bubble.

In this paper we consider a model of a scalar field with a global $O(3)$ 
symmetry and study the first-order phase transition at high temperature and
in the presence of gravity. We first show that, in addition to the ordinary
bubble of  which the center is true vacuum, the model in a curved spacetime
supports  the $O(3)$ bubble that includes a matter droplet in its core as
obtained in the same model in a flat spacetime \cite{Kim}. The formation of the
matter droplet at the core region of the new bubble is due to the winding
between the internal space and that of spatial rotations, so it can be
interpreted as a global monopole inside the bubble. Henceforth we will call
it ``{\it monopole-bubble}'' in order to distinguish this bubble solution
from the well-known {\it ordinary} bubble solution. This new bubble with a
global monopole has two bubble walls: one is the bubble wall which
distinguishes the false vacuum region outside the bubble from the true
vacuum region inside it, and the other is the inner bubble wall which
surrounds the false vacuum core of the global monopole. The long range tail
of global monopole energy density is inversely proportional to the square of 
the circumference radius and affects the spacetime structure inside the
bubble. Actually, it renders the spacetime between the inner and outer
walls similar to the spacetime around the ordinary global monopole
\cite{BV}: the spacetime has a repulsive nature, which indicates that the
core does not collapse into a black hole even at the Planck scale, and the
far region is a flat spacetime with a solid deficit angle. Although the total
energy of the ordinary global monopole is infinite, in the present case it
is finite since the outer wall plays an important role of cutoff. 

The value of the Euclidean action needed to generate a bubble with a matter
droplet is larger than  that of an ordinary clean bubble; however, we observe
that the decay rate for this monopole-bubble computed on the basis of the
exponential formula can be quite considerable in comparison with that of the
ordinary bubble for a certain range of the coefficients of the scalar potentials.
Once bubbles are nucleated at high temperature in a curved spacetime, they start
to expand because of some combination of processes when the environment keeps the
temperature high \cite{Ste} or the recovery of zero-temperature classical
dynamics by the expansion of the background universe. We investigate the
motion of the monopole-bubble by solving the coupled time-dependent field
equations, neglecting the process of temperature changing. The outer bubble
wall of the monopole-bubble grows similar to that of the ordinary bubble.
The global monopole at the center of the bubble is stable when the phase
transition scale is lower than the Planck scale. Moreover, this
monopole-bubble also shares the same possibility of defect inflation at the
Planck scale according to the arguments in Ref.\cite{Vil,SSTM,Sak}.

The paper is organized as follows. In Sec. II we introduce the model and
illustrate the nucleation of new $O(3)$ bubble solutions. In Sec. III we 
present the evolution of bubbles due to the gravitational force  and
briefly discuss the matter of defect inflation at the Planck scale at the
core of the global monopole.  Some concluding remarks are made in Sec. IV.
In this paper we use the units $c=\hbar=k_B=1$. 

\setcounter{section}{2}
\setcounter{equation}{0}
\bce\section*{\large\bf II. Nucleation of Monopole-Bubbles}
\ece
\indent\indent
Transition to the true vacuum state by quantum tunneling occurs through the
nucleation of bubbles of the energetically favored phase. The nucleation rate
per unit volume is $\Gamma=Ae^{-B}$, where $B$ is evaluated by the Euclidean
tunneling action and $A$ is a prefactor which has units of energy to the fourth
power. In this section we will consider a model with global $O(3)$ symmetry and
calculate the nucleation rate of bubbles at high temperature, specifically $B$
of the model in a curved spacetime.

\bce\subsection*{A. Euclidean Solutions}
\ece

We begin with the action of an $O(3)$ symmetric scalar multiplet in the 
presence of Einstein gravity. Since the quantum statistics of bosons at
finite temperature is formulated in the imaginary-time method which is
described by the Euclidean theory with fields periodic in the Euclidean time
with period $\beta=1/T$, the Euclidean action at finite temperature is given by
\be\label{eac}
S=\int^{\beta}_{0}dt_E\thint\!\sqrt{g}\;\biggl\{-\frac{1}{16\pi
G}R +\frac{1}{2}\ggtmn\pbm\ophi\pbn\ophi+V(\phi)\biggr\},
\ee
where 
$\phi^{a}=\hat{\phi}^{a}\phi$ is an 
$O(3)$-symmetric isovector $(a=1,2,3)$ with $\phi=\sqrt{\phi^{a}\phi^{a}}$.

The equations of motion are
\be\label{eqsc}
\frac{1}{\sqrt{g}}\pbm(\sqrt{g}\ggtmn\pbn\ophi)=-\frac{\partial(-V)}{
\partial\ophi}
\ee
\be\label{ein}
G_{\mu\nu}=8\pi G \tbmn ,
\ee
where the energy-momentum tensor is
\be
\tbmn=\pbm\ophi\pbn\ophi-\ggbmn\left(\frac{1}{2}\ggtrs\pbr\ophi\pbs\ophi+V
\right).
\ee
Here we bring up the situation that the temperature is much larger than the
inverse of the bubble radius, where the transition is described by
time-independent fields. Both scales of symmetry breaking and temperature are
usually assumed to be lower than the Planck scale $M_{Pl}=1/\sqrt{G}$; however,
sometimes the supermassive scale when the transition scale is comparable with
the Planck scale will also be considered when we examine the inner structure  of
bubbles in the very early universe.

Although there is no rigorous proof in a curved spacetime when the $O(3)$ 
symmetric solutions with  respect to spatial coordinates saturate the
solutions with the lower Euclidean action, we choose to consider the bubbles
which possess spherical symmetry. The form of the metric compatible with
$O(3)$ symmetry can be written as  \be\label{eme}
ds^{2}=\Bigl(1-\frac{2GM(r)}{r}\Bigr)e^{2\delta(r)}dt_E^{2}+\Bigl(1-
\frac{2GM(r)}{r}\Bigr)^{-1}dr^{2}+r^{2}(d\theta^{2}+\sin^{2}\theta d\varphi^{2}).
\ee
Using these coordinates the scalar field takes the following form,
\bea\label{sc}
\ophi&=&\hphi(\theta,\varphi)\phi(r)\nonumber\\
&=&(\sin n\theta\cos m\varphi,\;\sin n\theta\sin m\varphi,\;\cos n\theta)
\,\phi(r),
\eea
where the spherical symmetry and the regularity at the origin allow only 
two cases: $n=0$, and $n=m=1$.

Writing down the equations of motion by use of Eqs.(\ref{eme}) and (\ref{sc}),
we have three independent field equations:
\be\label{eq1}
\Bigl(1-\frac{2GM}{r}\Bigr)
\frac{d^{2}\phi}{dr^{2}}+
\Bigl(1-\frac{2GM}{r}\Bigr)
\frac{d}{dr}\ln\biggl(r^{2}e^{\delta}\Bigl(1-
\frac{2GM}{r}\Bigr)\biggr)\frac{d\phi}{dr}-\frac{2\delta_{n1}}{r^{2}}
\phi=\frac{dV}{d\phi},
\ee
\be\label{eq2}
\frac{1}{r}\frac{d\delta}{dr}=4\pi G\biggl(\frac{d\phi}{dr}\biggr)^{2},
\ee
\be\label{eq3}
\frac{1}{r^{2}}\frac{dGM}{dr}=4\pi G\biggl[\frac{1}{2}\Bigl(1-\frac{2GM}{r}
\Bigr)\biggl(\frac{d\phi}{dr}\biggr)^{2}+\frac{\delta_{n1}}{r^{2}}\phi^{2}
+V\biggr],
\ee
where $\delta_{n1}$ denotes the Kronecker delta. For the actual calculation, let
us choose a sixth-order scalar potential such as
\be\label{pot}
V(\phi)=\frac{\lambda}{v^{2}}(\phi^{2}+\alpha v^{2})(\phi^{2}-v^{2})^{2}
~~ {\rm with} ~~ 0<\alpha<1/2,
\ee
which describes the potential in a broken phase. Although we choose a 
specific shape of the scalar potential, our argument in the following does not
depend on the detailed form of the scalar potential and the  existence of the
new bubble solution with $n=1$ is guaranteed under any potential as long as it
includes one false vacuum and one true vacuum. Here we only consider the
transition from a symmetric vacuum to the broken vacuum, {\it i.e.} from de
Sitter spacetime with the horizon $H^{-1}\equiv(8\pi G V(0)/3)^{-{1\over2}}$ to
Minkowski spacetime.

The boundary conditions for nonsingular solutions of Eqs. (\ref{eq1}), (\ref{eq2})
and (\ref{eq3}) are
\bea\label{bc0}
\frac{d\phi}{dr}(r=0)=0 \;\mbox{for}\; n=0\; &\mbox{or}&
\phi(r=0)=0\;\mbox{for}\; n=1,\nonumber\\
\phi(r\rightarrow H^{-1})=0,&\mbox{and}& M(r=0)=0,
\eea
and we choose the normalization of $t$ by setting $\delta(r\rightarrow 
H^{-1})=0$. The scalar field rapidly approaches the false vacuum for large $r$
when the radius of the bubble is smaller than the de Sitter horizon $H^{-1}$.
The above is our main interest; however, we will also comment on the cases in
which the radius of the bubble is comparable to or larger than the de Sitter
horizon.

To find the behavior near the origin, we expand the variables by a power 
series, finding
\bea
\phi(r)&\approx&\phi_{{\rm esc}}
-\frac{1}{6}\frac{dV}{d\phi}\biggr|_{\phi_{{\rm esc}}}r^{2}
,~~\phi_{{\rm esc}}\equiv\phi(0)
\nonumber\\
\delta(r)&\approx&\delta_{{\rm esc}}
+\frac{\pi}{9}G\biggl(\frac{dV}{d\phi}
\biggr|_{\phi_{{\rm esc}}}\biggr)^{2}r^{4}
,~~\delta_{{\rm esc}}\equiv\delta(0)\\
GM(r)&\approx&\frac{4}{3}\pi GV(\phi_{{\rm esc}})r^{3}\nonumber
\eea
for the $n=0$ configuration, and 
\bea\label{bc1}
\phi(r)&\approx&\phi_{0}r\Bigl[1-\bigl(\frac{1}{2}\lambda(2-\alpha)
-\pi G\phi^{2}_{0})\bigr)r^{2}\Bigr]
,~~\phi_0\equiv{d\phi\over dr}(0)
\nonumber\\
\delta(r)&\approx&\delta_{0}+2\pi G\phi^{2}_{0}r^{2}
,~~\delta_{{\rm esc}}\equiv\delta(0)\\
GM(r)&\approx&2\pi G\phi_{0}^{2}r^{3}\nonumber
\eea
for the $n=1$ configuration. The constants $(\phi_{{\rm esc}},\delta_{{\rm
esc}})$ and $(\phi_{0}, \delta_{0})$ are determined by the proper behavior of the
fields for large $r$. If the transition scale
$m_{H}=\sqrt{4\lambda(3+2\alpha)}v$ is smaller than the Planck scale and
the size of the bubble is smaller than that of the de Sitter horizon $H^{-1}$, we
may meet no coordinate singularity inside the bubble. Therefore, the scalar
amplitude $\phi(r)$ and a metric function $\delta(r)$ approach their boundary
values exponentially in their asymptotic region, and Eq.(\ref{eq3}) says
\be
GM(r)\approx\frac{4}{3}\pi G\lambda\alpha v^{4}r^{3},
\ee
since the first two terms in the right-hand side of Eq.(\ref{eq3}) are 
negligible and the last cosmological constant term is dominant at the false
vacuum region outside the bubble. Figure 1 shows that $\delta(r)$ and $GM(r)$
are monotonically increasing functions of $r$; however, the behavior of the
scalar field $\phi$ near the origin depends on whether
$n=0$ or $n=1$.

\vspace{3mm}

\setlength{\unitlength}{0.240900pt}
\ifx\plotpoint\undefined\newsavebox{\plotpoint}\fi
\sbox{\plotpoint}{\rule[-0.200pt]{0.400pt}{0.400pt}}%


\vspace{11mm}

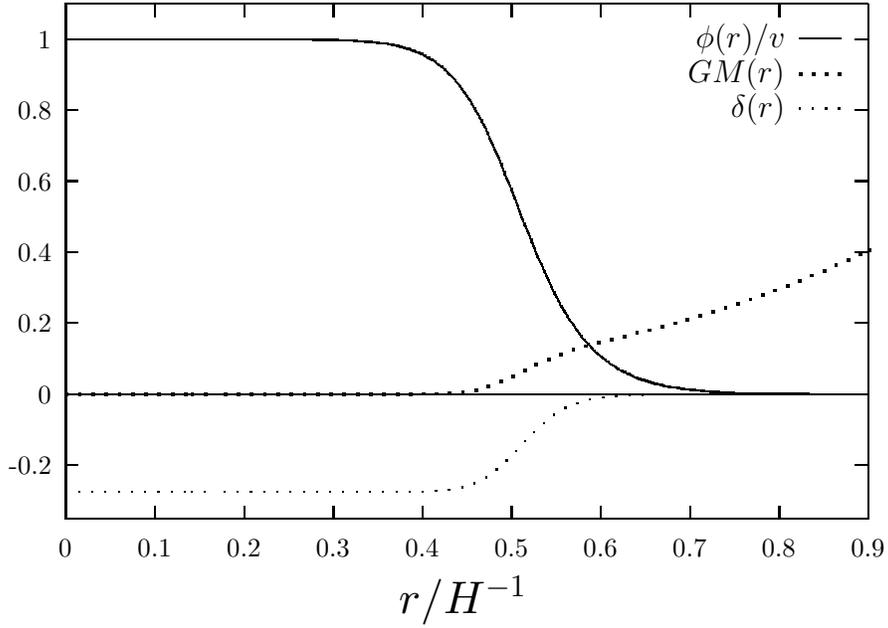
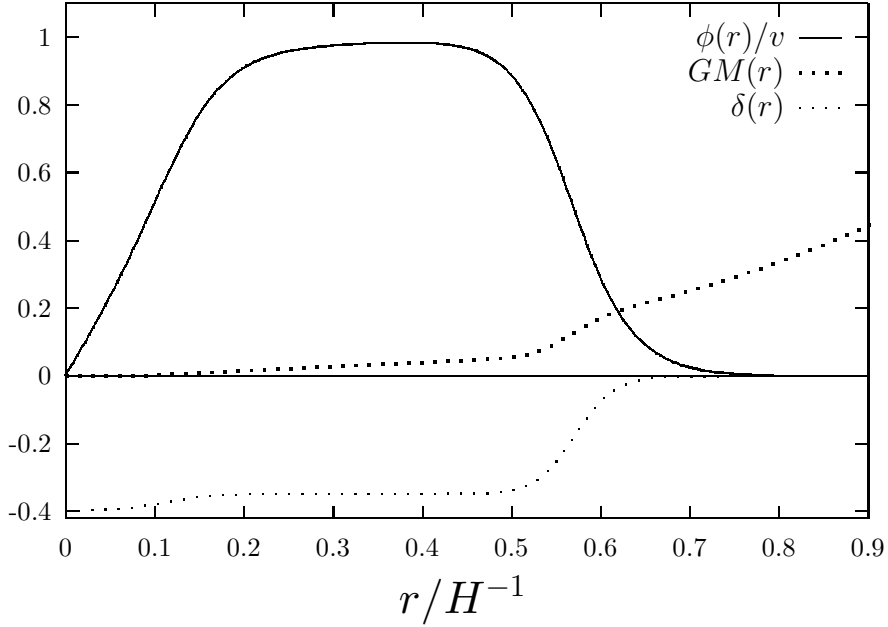
\noindent {\small Figure 1: A plot of a bubble solution for $\lambda=1$,
$\alpha=0.1$  and $v/M_{Pl}=0.1$. The solid, dotted and dashed lines correspond
to $\phi(r)$, $\delta(r)$ and $GM(r)$, respectively. (a) $n=0$, (b) $n=1$.}

\setlength{\unitlength}{0.240900pt}
\ifx\plotpoint\undefined\newsavebox{\plotpoint}\fi
\sbox{\plotpoint}{\rule[-0.200pt]{0.400pt}{0.400pt}}%
%

\vspace{16mm}

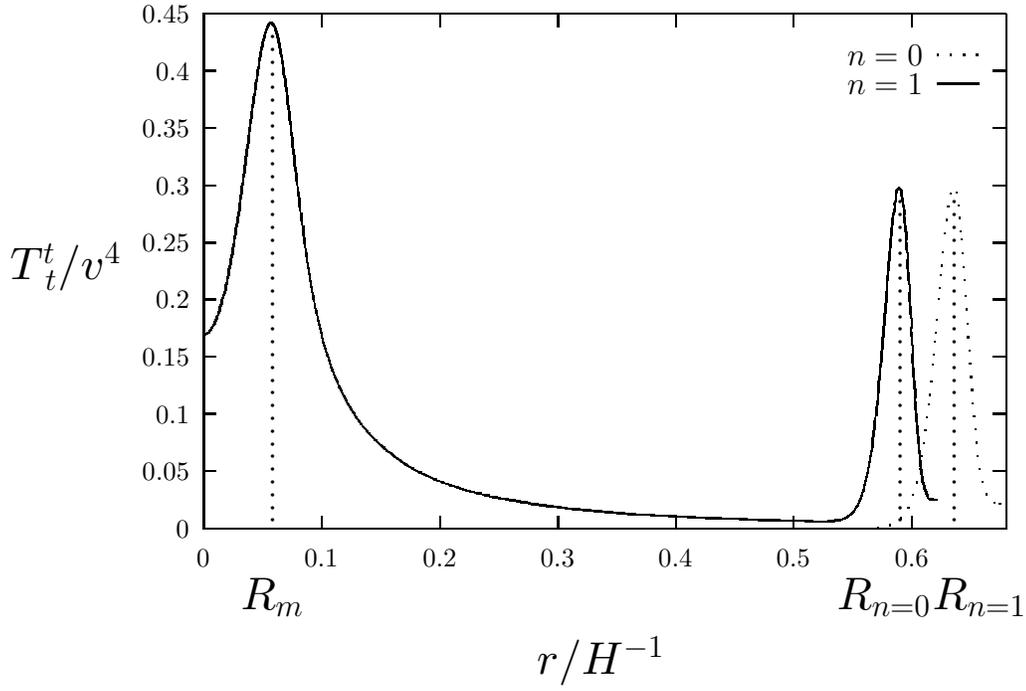
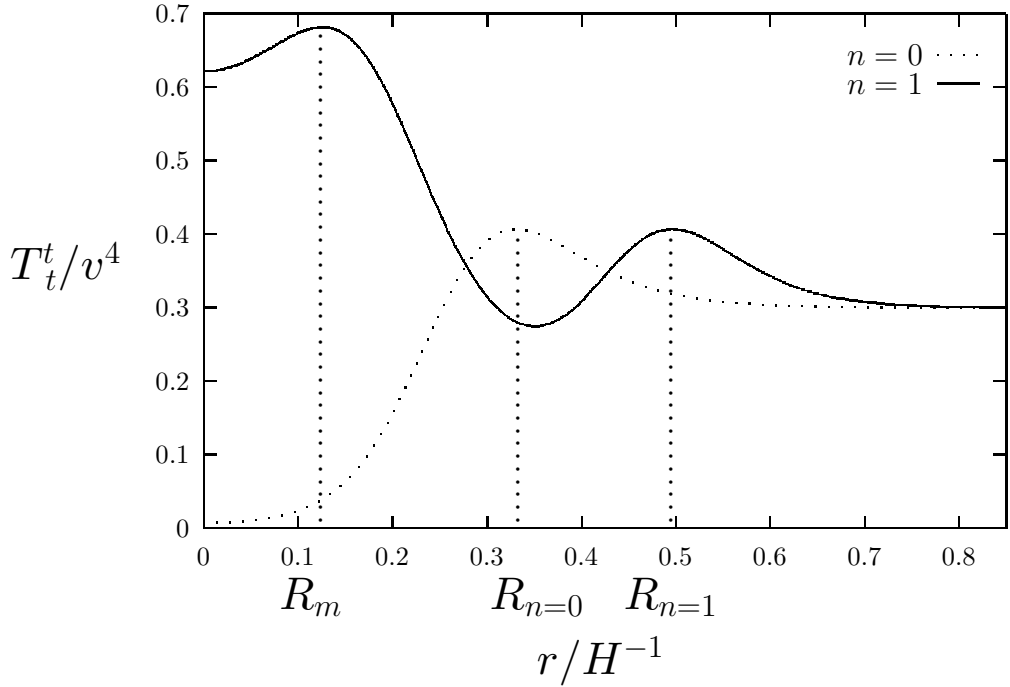
\noindent {\small Figure 2: $T^{t}_{\;t}$ profiles for fixed $\lambda=1$ and
$v/M_{Pl}=0.1$. (a) thick-wall bubble of $\alpha=0.3$, (b) thin-wall
bubble of $\alpha=0.01$. The dotted and solid lines correspond to an $n=0$
bubble and an $n=1$ bubble, respectively.}

\newpage

As shown in Fig. 2, the $n=0$ solution is the well-known vacuum bubble solution
where the energy is accumulated at the bubble wall. Although there remains some
matter inside the thick-wall bubble (see the dotted line in Fig. 2-(b)),
contrary to the case of the thin-wall bubble (see the dotted line in Fig.
2-(a)), such matter does not form an aggregate inside the bubble. The energy
density of $n=1$ bubbles,
\be\label{eng}
T^{t}_{\;t}=\frac{1}{2}\Bigl(1-\frac{2GM}{r}\Bigr)\biggl(
\frac{d\phi}{dr}\biggr)^{2}+\delta_{n1}\frac{\phi^{2}}{r^{2}}+V,
\ee
is expressed by solid lines in Fig. 2. We easily read that a matter droplet is
formed at the center of the $n=1$ bubble due to the nontrivial local mapping
between internal $O(3)$ symmetry and spatial $O(3)$ symmetry. From the boundary
conditions of the scalar field in Eq.(\ref{bc0}) together with the ansatz
(\ref{sc}), we can interpret this matter droplet as a global monopole. In a flat
spacetime of dimension more than two, there is a no-go theorem that says the
scalar fields described by the standard relativistic form of the Lagrangian do
not support non-trivial static soliton solutions of finite energy \cite{DH}.
Therefore, when we consider global vortices or global monopoles in the presence
of gravity \cite{Gre,BV}, the introduction of a cutoff scale, for example, the 
horizon length, provides us a way to control the divergent quantities \cite{HP}.
However, the global monopole created inside the $n=1$ bubble is a finite energy
static configuration since the long-range tail of the global monopole is tamed
by the outer bubble wall.

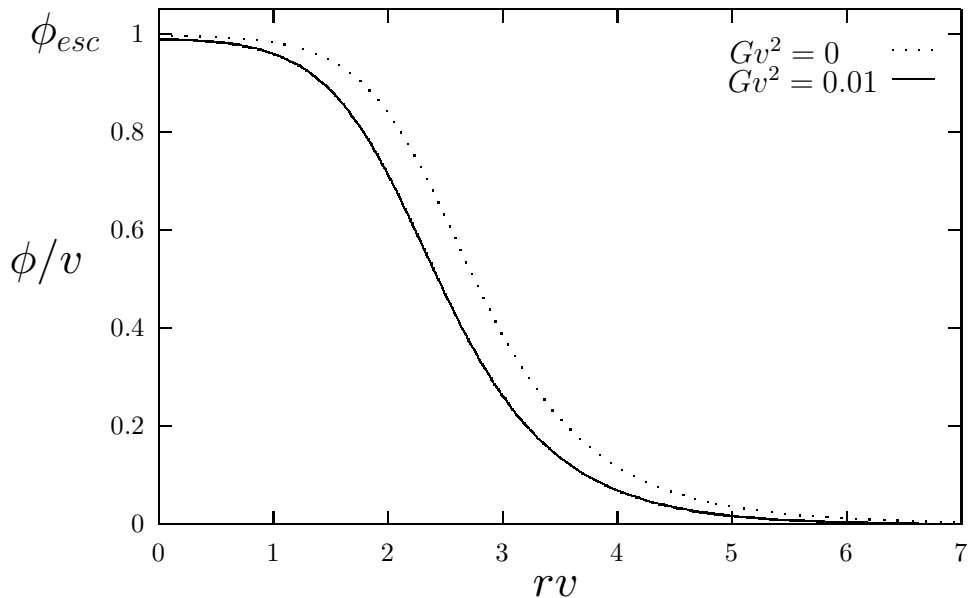
\vspace{8mm}

\setlength{\unitlength}{0.240900pt}
\ifx\plotpoint\undefined\newsavebox{\plotpoint}\fi
\sbox{\plotpoint}{\rule[-0.200pt]{0.400pt}{0.400pt}}%


\vspace{10mm}

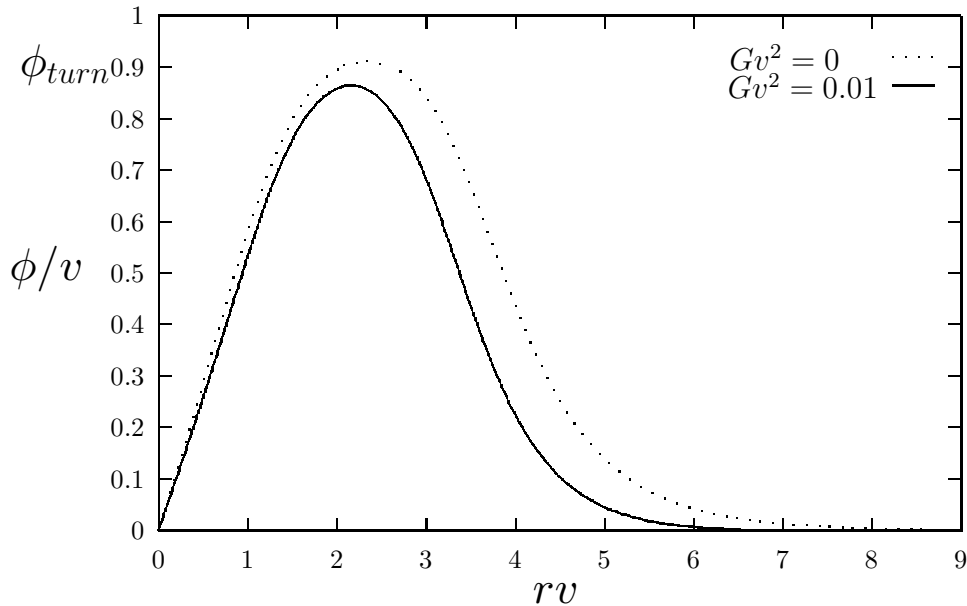
\noindent {\small Figure 3: A plot of bubble solutions with and without gravity
coupling for $\lambda=1$ and $\alpha=0.25$. The solid line denotes the
bubble with gravity and the dotted line without gravity. (a) $n=0$, (b)
$n=1$.}

\vspace{5mm}

The features described above do not depend on whether the bubbles lie in a 
flat spacetime or a curved spacetime; however, other characteristics due to
gravity are also worth noting. We can read the gravitational effects on the
shapes of bubbles in the weak gravity limit as follows. If we think of $r$ as a
time and $\phi$ as a position of a hypothetical particle in a one-dimensional
motion, the scalar field equation (\ref{eq1}) can be interpreted as the
Newton equation for a particle of variable mass
$\displaystyle{\Bigl(1-\frac{2GM}{r}\Bigr)}$ subject to several forces: the
first one is the conservative force from $-V(\phi)$, the second  one a friction
with time-dependent coefficient
$\displaystyle{\Bigl(1-\frac{2GM}{r}\Bigr)\frac{d}{dr}\ln\Bigl(
r^{2}e^{\delta}\bigl(1-\frac{2GM}{r})\Bigr)}$, and the last one only for  $n=1$
is a time-dependent repulsion whose strength is extremely large at $r=0$. 
Although it is difficult to prove analytically the existence of solutions in the
presence of gravity, we can understand how and what kind of bubble solutions are
supported by use of the terminology of Newtonian mechanics as was done in the
flat spacetime case \cite{KKK}. Let us consider the returning motion of a
hypothetical particle both for $n=0$ and $n=1$ solutions, {\it i.e.}, from
$\phi(r=0)=\phi_{{\rm esc}}$ for the $n=0$ solution (or $\phi(r=r_{{\rm
turn}})=\phi_{{\rm turn}}$ for the $n=1$ solution) to $\phi(r=\infty)=0$. Since
$GM(r)$ is an increasing function of $r$, the variable mass of a hypothetical
particle $\displaystyle{\Bigl(1-\frac{2GM}{r}\Bigr)}$ decreases as time $r$
elapses. Furthermore, the gravitational effect due to the sum of $\delta(r)$ and
$GM(r)$ decreases the time-dependent coefficient at the starting point for the 
$n=0$ solution and at the turning point for the $n=1$ solution. Both effects
involve the decrement of energy gained during the rolling of a hypothetical
particle down to the minimum point of the  effective potential $-V(\phi)$,
hence $\phi_{{\rm esc}}$ (or $\phi_{{\rm turn}}$) should  become small after the
inclusion of gravity (see Fig. 3). For an $n=0$ bubble, it makes the radius of
the bubble smaller. For the behavior of the $n=1$ bubble, the hypothetical
particle reaches the returning point, $\phi_{{\rm turn}}$, at a time earlier
than in the flat case. In order to reach the smaller turning point $\phi_{\rm
turn}$ in spite of the larger acceleration for a given $\phi_{0}$ and for small
$r$ as shown in Eq.(\ref{bc1}), the initial velocity $\phi_{0}$ of the particle
for $n=1$ bubbles should be smaller than the value in a flat spacetime. Hence,
the radius of the bubble becomes small, although the size of the matter droplet
is increased by the gravity. The contraction of the outer walls of both
$n=0,\;1$ bubbles can be understood through the attractive nature of gravity;
however, the explanation for the larger matter droplet will be given later by
mentioning the repulsive nature of gravity on the matter  core.   Looking back
at Fig. 2, we can compare the radius of an $n=0$ bubble and of an $n=1$ bubble.
In a flat spacetime the radius of an $n=0$ bubble is always larger than that of
an $n=1$ bubble \cite{Kim}, and it is also true for a thick-wall bubble in a
curved spacetime (see Fig. 2-(b)). For a thin-wall bubble in a curved spacetime
where the global monopole stretches a long-range tail; however, the radius of
an $n=1$ bubble becomes smaller than that of an $n=0$ bubble. This fact will be
justified by our analytic discussion under the thin-wall approximation in Sec.
III.

For an $n=1$ bubble, let us look at the spacetime structure in the 
neighborhood  of the point $r=r_{{\rm turn}}$ where $\phi$ takes the
maximum value $\phi_{{\rm turn}}$, {\it i.e.}, 
$\displaystyle{\frac{d\phi}{dr}\biggr|_{r=r_{{\rm turn}}}}=0$. Equations
(\ref{eq2}) and (\ref{eq3}) around $r=r_{{\rm turn}}=0$ have approximate
solutions such as \bea \delta(r)&\approx&\delta_{{\rm turn}}\label{dso}\\
GM(r)&\approx&GM_{{\rm turn}}+4\pi G\phi^{2}_{{\rm turn}}r+\frac{4}{3}\pi 
GV(\phi_{{\rm turn}}) r^{3}
\label{mso}.\eea
The constants $\delta_{{\rm turn}}$ and $M_{{\rm turn}}$ are fixed by
\bea
\delta_{{\rm turn}}&\approx&-4\pi G \int^{H^{-1}}_{r_{{\rm turn}}}
dr\, r \biggl(\frac{d\phi}{dr}\biggr)^{2}\label{dtu}\\
GM_{{\rm turn}}\!\!&\approx&\!\!4\pi
G\!\int^{r_{{\rm turn}}}_{0}\!\!\!dr\,r^{2}\biggl\{\frac{1}{2}
\Bigl(1-\frac{2GM}{r}\Bigr)\biggl(\frac{d\phi}{dr}\biggr)^{2}\!
+\frac{\phi^{2}-\phi^{2}_{{\rm turn}}}{r^{2}}
+\bigl(V-V(\phi_{{\rm turn}})\bigr)\biggr\}.
\label{mtu} \eea
The above integrals are estimated as $\delta_{{\rm turn}}\sim 4\pi G
\phi^{2}_{{\rm turn}}$ and $|GM_{{\rm turn}}|\sim|Gm_{H}|$.

If we consider an $n=1$ bubble with a thin wall when the phase transition 
scale $m_{H}= \sqrt{4\lambda(3+2\alpha)}v$ is much smaller than the Planck
scale $M_{Pl}$, the first and the third terms in Eq.(\ref{mso}) can be
neglected since $V(\phi_{{\rm turn}})\approx 0$ and $vr_{{\rm turn}}\gg 1$.
Substituting the above results into Eq.(\ref{eme}) and rescaling the
variables $t$ and $r$ as \bea
t&\rightarrow&(1-8\pi G\phi^{2}_{{\rm turn}})^{-\frac{1}{2}}e^{-\delta_{{\rm turn}}}t
\label{tres}\\
r&\rightarrow&(1-8\pi G\phi^{2}_{{\rm turn}})^{\frac{1}{2}}r,
\label{rres}
\eea
we obtain a metric after a Wick rotation, which describes the region inside 
the outer wall but outside the global monopole:
\be\label{flat}
ds^{2}=-dt^{2}+dr^{2}+r^{2}(1-8\pi G\phi_{{\rm turn}}^{2})(d\theta^{2}+\sin^2\theta\,
d\varphi^{2}).
\ee
Although the actual metric is not completely flat due to the additional 
small terms in Eq.(\ref{mso}), the observer inside the $n=1$ bubble feels
no gravitational force exerted by the global monopole apart from the tiny
effects from the monopole core and  the energy difference between the true
vacuum $v$ and the maximum value of the $n=1$ bubble $\phi_{{\rm turn}}$,
$V(\phi_{{\rm turn}})-V(v)$. This phenomenon can be explained by a
Newtonian gravitational potential. The radial component of tension
$-T^{r}_{\;r}$ also has a long range term such as \be
T^{r}_{\;r}=-\frac{1}{2}\Bigl(1-\frac{2GM}{r}\Bigr)\Bigl(
\frac{d\phi}{dr}\Bigr)^{2}+\delta_{n1}\frac{\phi^{2}}{r^{2}}+V,
\ee
which cancels the energy density $T^{t}_{\;t}$ in (\ref{eng}) in the 
Newtonian limit of the Einstein equations: $\nabla^{2}\Phi=8\pi G(T^{t}_{\;t}-
T^{r}_{\;r}){\approx}0$ at $r\approx r_{{\rm turn}}$. However, since the
metric (\ref{flat}) describes a space with a deficit solid angle, if we
consider a light signal propagating from a source to an observer, the
observer  inside the $n=1$ bubble must notice the light bending due to the
deficit solid angle $\Delta=8\pi Gv^{2}$, as is the case of a straight cosmic
string. Thus a rough evaluation gives the angular separation
$\delta\varphi\sim 8\pi Gv^{2}\sim$few $arcsec$ at a typical grand
unification scale $v\sim10^{16}GeV$, which can be observable.

For the phase transition in the supermassive scale, the absolute value of 
$M_{{\rm turn}}$ in Eq.(\ref{mtu}) is of the order of the Planck scale,
$|M_{{\rm turn}}|\sim M_{Pl}$, and then the first term in Eq.(\ref{mso})
becomes considerably large. Let us discuss the structure of the spacetime
in this case. Assuming the outer wall is extremely thin, {\it i.e.}, the
third term in the right hand side of Eq.(\ref{mso}) is negligible inside
the bubble, we obtain a metric around $\phi(r)\sim\phi_{{\rm turn}}$ as
\be\label{mout} ds^{2}=-\Bigl(1-8\pi Gv^{2}
-\frac{2GM_{{\rm turn}}}{r}\Bigr)e^{2\delta_{{\rm turn}}}dt^{2}+ 
\Bigl(1-8\pi Gv^{2} -\frac{2GM_{{\rm turn}}}{r}\Bigr)^{-1}dr^{2}+r^{2}
(d\theta^{2}+\sin^{2}\theta d\varphi^{2}).
\ee
Here we make a crude assumption that the region between the inner wall and 
the outer wall ($R_{m}\leq r\leq R_{n=1}$, see Fig. 2) is described by the
above metric in Eq.(\ref{mout}), and the region inside the inner wall ($r\leq
R_{m}$) is approximated to be de Sitter spacetime:
\be\label{dS}
ds^{2}=-\Bigl(1-\frac{8}{3}\pi G V(0)r^{2}\Bigr)e^{2\delta_{0}}
dt^{2}+\Bigl(1-\frac{8}{3}\pi G V(0)r^{2}\Bigr)^{-1}dr^{2}
+r^{2}(d\theta^{2}+\sin^2\theta d\varphi^{2}).
\ee
Equation (\ref{eq2}) says that the change of $\delta(r)$ can be neglected 
inside the monopole, which is supported by Fig. 1(b), and thus we take
$\delta_{0}\approx\delta_{{\rm turn}}$ under our assumption. From the 
continuity of the metric and its first derivative with respect to $r$, we
estimate the size of the global monopole (or equivalently the position of
the inside bubble wall) as $R_{m}=1/\sqrt{4\lambda(3+2\alpha)}v$ and
$M_{{\rm turn}} = -\sqrt{4\lambda(3+2\alpha)} v<0$. It is confirmed by
envisaging the integral in (\ref{mtu}): if we divide the integration domain
into ($0,R_{m}$) and  ($R_{m},r_{{\rm turn}}$), and then substitute the
values of the scalar amplitude, {\it i.e.}, $\phi=0$ in $(0,R_{m})$ and
$\phi=\phi_{{\rm turn}}$ in ($R_{m}, r_{{\rm turn}}$), the integral of the
core region has only the negative contribution to $GM_{{\rm turn}}$. This
result is consistent with the known result that the global monopole inside
the $n=1$ bubble does not form a black hole even at the Planck scale
\cite{BV}. It can also be checked by the radial motion of a test particle
governed by the geodesic equation
\be\label{acc}
\frac{d^{2}r}{d\tau^{2}}=\frac{d}{dr}\biggl(\frac{GM}{r}\biggr)-
\Bigl(1-\frac{2GM}{r}\Bigr)\frac{d\delta}{dr}, 
\ee
where $\tau$ is the proper time of a test particle. Equation (\ref{acc}), 
together with (\ref{bc1}), tells us that the acceleration increases in
proportion to the radius around the center of the global monopole. Although it is
a weak gravity case ($v/M_{Pl}=0.1$), Fig. 4 shows a typical example consistent
with the above argument. Here a question arises: what is the structure of a
spacetime manifold which is formed when the deficit solid angle is equal to or
greater than $4\pi$? In the case of local cosmic strings, when the deficit angle
is equal to or greater than $2\pi$, a possible two-dimensional spatial manifold
is described by cylinder or two sphere \cite{Cosm}. However, it is an open
question for the global monopole.

\vspace{5mm}

\setlength{\unitlength}{0.240900pt}
\ifx\plotpoint\undefined\newsavebox{\plotpoint}\fi


\vspace{10mm}

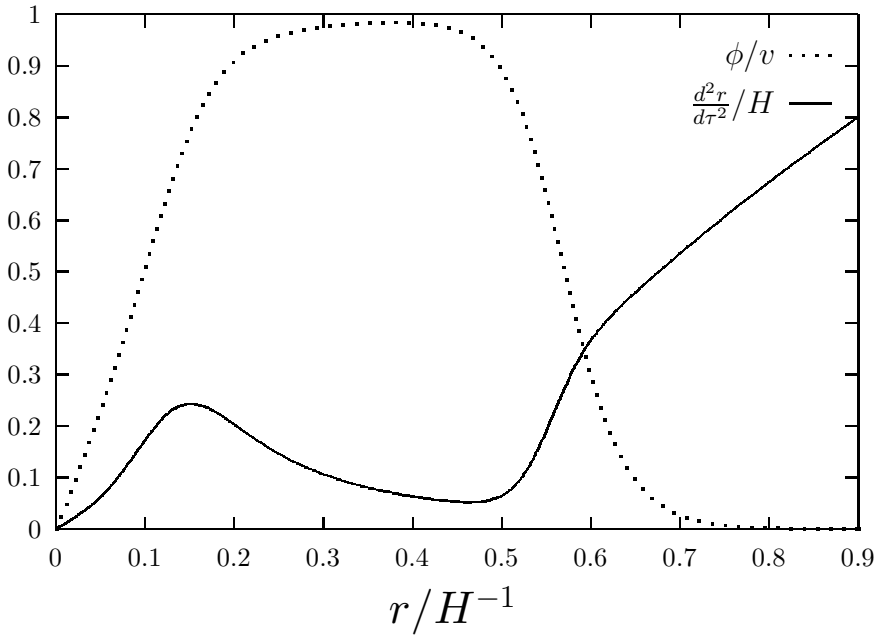
\noindent {\small Figure 4: The acceleration of a test particle
$d^{2}r/d\tau^{2}$ versus the radius $r$ for $\lambda=1$ and $\alpha=0.1$. The
acceleration denoted by the solid line is always positive and increases at the
core of the global  monopole for a weak gravity case ($v/M_{Pl}=0.1$).}

\bce\subsection*{B. Nucleation Rate}
\ece

We now turn to the evaluation of the nucleation rate of bubbles; specifically
the values of the Euclidean action at two stationary points of $n=0$ ($B_{0}$)
and of $n=1$ ($B_{1}$) bubbles are of our interest. Suppose that there exists a
barrier between a local minimum of $B_{0}$ and $B_{1}$, and the maximum value of
$B$ between $B_{0}$ and $B_{1}$ can be set to be $B_{{\rm top}}$. When each
valley is so deep that the height of the barrier at the hilltop $B_{{\rm top}}$
evaluated from the ordinary bubble point is larger than the difference of two
local minima at the bottoms of valleys, {\it i.e.}, $B_{{\rm top}}-B_{1}\gg
B_{1}-B_{0}$, then the total decay rate from a metastable phase into a stable
phase per unit volume can be estimated to be the sum of the nucleation rate of
each bubble \be \Gamma=\Gamma^{(0)}+\Gamma^{(1)}.
\ee
It means that both $n=0$ and $n=1$ solutions give distinct decay
channels where each solution describes the nucleation of bubbles with the 
critical size. Moreover, if the tunneling action for each bubble is larger
than unity, the nucleation rate for the $n$-th bubble takes the exponential
form
\be\label{exfa}
\Gamma^{(n)}=A_{n}e^{-B_{n}}.
\ee
On the other hand, if the tunneling action is of the order of unity or smaller,
which corresponds to thick-wall bubbles in the high temperature limit, the
exponential formula (\ref{exfa}) is no longer valid. Here we take a heuristic
viewpoint and keep our analysis on the basis of the above formula in
Eq.(\ref{exfa}), notwithstanding the above possibility. When the first-order
phase transition is considered in a curved spacetime, the background spacetime
itself experiences time evolution, {\it e.g.}, inflation in the metastable
vacuum region and this expansion of spatial volume induces a sudden drop in
temperature. Therefore, the shapes of bubbles or equivalently the configurations
of the scalar field change, and hence the formula (\ref{exfa}) should not be
applied.  
Here we discuss
the probability to nucleate bubbles when the system is initially in the
metastable phase and the temperature change can be neglected \cite{Hog}. Under
this restriction $B_{n}$ can be approximated to be the values of action 
(\ref{eac}) for a given $n=0$ or $n=1$ bubble since we consider only the decay
from a spacetime with a positive cosmological constant, $V(0)>0$, into a
spacetime with zero cosmological constant, $V(v)=0$.

We plot a dimensionless value $B^{'}_{n}\equiv(T/v)B_{n}$ for various $v/M_{Pl}$
and $\alpha$ in Figs. 5 and 6, respectively. Figure 5 shows that  $B^{'}_{n}$
becomes small for large $v/M_{Pl}$, which implies that the materialization of
both $n=0$ and $n=1$ bubbles is more likely at higher energy scales. This result
is consistent with the fact that the gravitational contribution to the total
energy is always negative, as we have already seen in Fig. 3.

\vspace{7mm}

\setlength{\unitlength}{0.240900pt}
\ifx\plotpoint\undefined\newsavebox{\plotpoint}\fi
\sbox{\plotpoint}{\rule[-0.200pt]{0.400pt}{0.400pt}}%
\begin{picture}(1500,900)(0,0)
\font\gnuplot=cmr10 at 10pt
\gnuplot
\sbox{\plotpoint}{\rule[-0.200pt]{0.400pt}{0.400pt}}%

\put(0,480){\makebox(0,0)[1]{\shortstack{\Large $\frac{T}{v}B_{n}$}}}
\put(800,-50){\makebox(0,0){\Large $v/M_{Pl}$}}

\put(176.0,68.0){\rule[-0.200pt]{0.400pt}{194.888pt}}
\put(176.0,68.0){\rule[-0.200pt]{4.818pt}{0.400pt}}
\put(154,68){\makebox(0,0)[r]{10}}
\put(1416.0,68.0){\rule[-0.200pt]{4.818pt}{0.400pt}}
\put(176.0,184.0){\rule[-0.200pt]{4.818pt}{0.400pt}}
\put(154,184){\makebox(0,0)[r]{20}}
\put(1416.0,184.0){\rule[-0.200pt]{4.818pt}{0.400pt}}
\put(176.0,299.0){\rule[-0.200pt]{4.818pt}{0.400pt}}
\put(154,299){\makebox(0,0)[r]{30}}
\put(1416.0,299.0){\rule[-0.200pt]{4.818pt}{0.400pt}}
\put(176.0,415.0){\rule[-0.200pt]{4.818pt}{0.400pt}}
\put(154,415){\makebox(0,0)[r]{40}}
\put(1416.0,415.0){\rule[-0.200pt]{4.818pt}{0.400pt}}
\put(176.0,530.0){\rule[-0.200pt]{4.818pt}{0.400pt}}
\put(154,530){\makebox(0,0)[r]{50}}
\put(1416.0,530.0){\rule[-0.200pt]{4.818pt}{0.400pt}}
\put(176.0,646.0){\rule[-0.200pt]{4.818pt}{0.400pt}}
\put(154,646){\makebox(0,0)[r]{60}}
\put(1416.0,646.0){\rule[-0.200pt]{4.818pt}{0.400pt}}
\put(176.0,761.0){\rule[-0.200pt]{4.818pt}{0.400pt}}
\put(154,761){\makebox(0,0)[r]{70}}
\put(1416.0,761.0){\rule[-0.200pt]{4.818pt}{0.400pt}}
\put(176.0,877.0){\rule[-0.200pt]{4.818pt}{0.400pt}}
\put(154,877){\makebox(0,0)[r]{80}}
\put(1416.0,877.0){\rule[-0.200pt]{4.818pt}{0.400pt}}
\put(176.0,68.0){\rule[-0.200pt]{0.400pt}{4.818pt}}
\put(176,23){\makebox(0,0){0}}
\put(176.0,857.0){\rule[-0.200pt]{0.400pt}{4.818pt}}
\put(324.0,68.0){\rule[-0.200pt]{0.400pt}{4.818pt}}
\put(324,23){\makebox(0,0){0.02}}
\put(324.0,857.0){\rule[-0.200pt]{0.400pt}{4.818pt}}
\put(472.0,68.0){\rule[-0.200pt]{0.400pt}{4.818pt}}
\put(472,23){\makebox(0,0){0.04}}
\put(472.0,857.0){\rule[-0.200pt]{0.400pt}{4.818pt}}
\put(621.0,68.0){\rule[-0.200pt]{0.400pt}{4.818pt}}
\put(621,23){\makebox(0,0){0.06}}
\put(621.0,857.0){\rule[-0.200pt]{0.400pt}{4.818pt}}
\put(769.0,68.0){\rule[-0.200pt]{0.400pt}{4.818pt}}
\put(769,23){\makebox(0,0){0.08}}
\put(769.0,857.0){\rule[-0.200pt]{0.400pt}{4.818pt}}
\put(917.0,68.0){\rule[-0.200pt]{0.400pt}{4.818pt}}
\put(917,23){\makebox(0,0){0.1}}
\put(917.0,857.0){\rule[-0.200pt]{0.400pt}{4.818pt}}
\put(1065.0,68.0){\rule[-0.200pt]{0.400pt}{4.818pt}}
\put(1065,23){\makebox(0,0){0.12}}
\put(1065.0,857.0){\rule[-0.200pt]{0.400pt}{4.818pt}}
\put(1214.0,68.0){\rule[-0.200pt]{0.400pt}{4.818pt}}
\put(1214,23){\makebox(0,0){0.14}}
\put(1214.0,857.0){\rule[-0.200pt]{0.400pt}{4.818pt}}
\put(1362.0,68.0){\rule[-0.200pt]{0.400pt}{4.818pt}}
\put(1362,23){\makebox(0,0){0.16}}
\put(1362.0,857.0){\rule[-0.200pt]{0.400pt}{4.818pt}}
\put(176.0,68.0){\rule[-0.200pt]{303.534pt}{0.400pt}}
\put(1436.0,68.0){\rule[-0.200pt]{0.400pt}{194.888pt}}
\put(176.0,877.0){\rule[-0.200pt]{303.534pt}{0.400pt}}
\put(176.0,68.0){\rule[-0.200pt]{0.400pt}{194.888pt}}
\put(1306,812){\makebox(0,0)[r]{$n=0$}}
\multiput(1328,812)(20.756,0.000){4}{\usebox{\plotpoint}}
\put(176,373){\usebox{\plotpoint}}
\multiput(176,373)(20.738,-0.841){4}{\usebox{\plotpoint}}
\multiput(250,370)(20.688,-1.677){4}{\usebox{\plotpoint}}
\multiput(324,364)(20.522,-3.102){4}{\usebox{\plotpoint}}
\multiput(410,351)(20.377,-3.944){3}{\usebox{\plotpoint}}
\multiput(472,339)(20.242,-4.588){4}{\usebox{\plotpoint}}
\multiput(547,322)(20.048,-5.374){4}{\usebox{\plotpoint}}
\multiput(644,296)(19.957,-5.702){3}{\usebox{\plotpoint}}
\multiput(700,280)(19.880,-5.964){3}{\usebox{\plotpoint}}
\multiput(750,265)(19.831,-6.124){3}{\usebox{\plotpoint}}
\multiput(818,244)(19.907,-5.874){3}{\usebox{\plotpoint}}
\multiput(879,226)(19.937,-5.771){2}{\usebox{\plotpoint}}
\multiput(917,215)(19.850,-6.065){2}{\usebox{\plotpoint}}
\multiput(953,204)(19.942,-5.753){3}{\usebox{\plotpoint}}
\multiput(1005,189)(20.034,-5.426){2}{\usebox{\plotpoint}}
\multiput(1053,176)(20.097,-5.186){2}{\usebox{\plotpoint}}
\put(1104.02,162.66){\usebox{\plotpoint}}
\multiput(1114,160)(20.108,-5.144){2}{\usebox{\plotpoint}}
\multiput(1157,149)(20.136,-5.034){2}{\usebox{\plotpoint}}
\put(1204.58,137.04){\usebox{\plotpoint}}
\multiput(1224,132)(20.382,-3.920){2}{\usebox{\plotpoint}}
\multiput(1250,127)(20.197,-4.783){2}{\usebox{\plotpoint}}
\put(1305.75,114.05){\usebox{\plotpoint}}
\multiput(1324,110)(20.319,-4.233){2}{\usebox{\plotpoint}}
\put(1366.62,100.95){\usebox{\plotpoint}}
\put(1387.03,97.17){\usebox{\plotpoint}}
\put(1405,94){\usebox{\plotpoint}}
\put(1306,767){\makebox(0,0)[r]{$n=1$}}
\put(1328.0,767.0){\rule[-0.200pt]{15.899pt}{0.400pt}}
\put(176,836){\usebox{\plotpoint}}
\multiput(176.00,834.93)(8.168,-0.477){7}{\rule{6.020pt}{0.115pt}}
\multiput(176.00,835.17)(61.505,-5.000){2}{\rule{3.010pt}{0.400pt}}
\multiput(250.00,829.92)(2.354,-0.494){29}{\rule{1.950pt}{0.119pt}}
\multiput(250.00,830.17)(69.953,-16.000){2}{\rule{0.975pt}{0.400pt}}
\multiput(324.00,813.92)(1.396,-0.497){59}{\rule{1.210pt}{0.120pt}}
\multiput(324.00,814.17)(83.489,-31.000){2}{\rule{0.605pt}{0.400pt}}
\multiput(410.00,782.92)(1.075,-0.497){55}{\rule{0.955pt}{0.120pt}}
\multiput(410.00,783.17)(60.017,-29.000){2}{\rule{0.478pt}{0.400pt}}
\multiput(472.00,753.92)(0.917,-0.498){79}{\rule{0.832pt}{0.120pt}}
\multiput(472.00,754.17)(73.274,-41.000){2}{\rule{0.416pt}{0.400pt}}
\multiput(547.00,712.92)(0.783,-0.499){121}{\rule{0.726pt}{0.120pt}}
\multiput(547.00,713.17)(95.494,-62.000){2}{\rule{0.363pt}{0.400pt}}
\multiput(644.00,650.92)(0.758,-0.498){71}{\rule{0.705pt}{0.120pt}}
\multiput(644.00,651.17)(54.536,-37.000){2}{\rule{0.353pt}{0.400pt}}
\multiput(700.00,613.92)(0.715,-0.498){67}{\rule{0.671pt}{0.120pt}}
\multiput(700.00,614.17)(48.606,-35.000){2}{\rule{0.336pt}{0.400pt}}
\multiput(750.00,578.92)(0.709,-0.498){93}{\rule{0.667pt}{0.120pt}}
\multiput(750.00,579.17)(66.616,-48.000){2}{\rule{0.333pt}{0.400pt}}
\multiput(818.00,530.92)(0.710,-0.498){83}{\rule{0.667pt}{0.120pt}}
\multiput(818.00,531.17)(59.615,-43.000){2}{\rule{0.334pt}{0.400pt}}
\multiput(879.00,487.92)(0.732,-0.497){49}{\rule{0.685pt}{0.120pt}}
\multiput(879.00,488.17)(36.579,-26.000){2}{\rule{0.342pt}{0.400pt}}
\multiput(917.00,461.92)(0.752,-0.496){45}{\rule{0.700pt}{0.120pt}}
\multiput(917.00,462.17)(34.547,-24.000){2}{\rule{0.350pt}{0.400pt}}
\multiput(953.00,437.92)(0.766,-0.498){65}{\rule{0.712pt}{0.120pt}}
\multiput(953.00,438.17)(50.523,-34.000){2}{\rule{0.356pt}{0.400pt}}
\multiput(1005.00,403.92)(0.776,-0.497){59}{\rule{0.719pt}{0.120pt}}
\multiput(1005.00,404.17)(46.507,-31.000){2}{\rule{0.360pt}{0.400pt}}
\multiput(1053.00,372.92)(0.820,-0.495){35}{\rule{0.753pt}{0.119pt}}
\multiput(1053.00,373.17)(29.438,-19.000){2}{\rule{0.376pt}{0.400pt}}
\multiput(1084.00,353.92)(0.838,-0.495){33}{\rule{0.767pt}{0.119pt}}
\multiput(1084.00,354.17)(28.409,-18.000){2}{\rule{0.383pt}{0.400pt}}
\multiput(1114.00,335.92)(0.864,-0.497){47}{\rule{0.788pt}{0.120pt}}
\multiput(1114.00,336.17)(41.364,-25.000){2}{\rule{0.394pt}{0.400pt}}
\multiput(1157.00,310.92)(0.874,-0.496){43}{\rule{0.796pt}{0.120pt}}
\multiput(1157.00,311.17)(38.349,-23.000){2}{\rule{0.398pt}{0.400pt}}
\multiput(1197.00,287.92)(0.974,-0.494){25}{\rule{0.871pt}{0.119pt}}
\multiput(1197.00,288.17)(25.191,-14.000){2}{\rule{0.436pt}{0.400pt}}
\multiput(1224.00,273.92)(0.938,-0.494){25}{\rule{0.843pt}{0.119pt}}
\multiput(1224.00,274.17)(24.251,-14.000){2}{\rule{0.421pt}{0.400pt}}
\multiput(1250.00,259.92)(1.065,-0.495){33}{\rule{0.944pt}{0.119pt}}
\multiput(1250.00,260.17)(36.040,-18.000){2}{\rule{0.472pt}{0.400pt}}
\multiput(1288.00,241.92)(0.954,-0.495){35}{\rule{0.858pt}{0.119pt}}
\multiput(1288.00,242.17)(34.219,-19.000){2}{\rule{0.429pt}{0.400pt}}
\multiput(1324.00,222.92)(1.225,-0.491){17}{\rule{1.060pt}{0.118pt}}
\multiput(1324.00,223.17)(21.800,-10.000){2}{\rule{0.530pt}{0.400pt}}
\multiput(1348.00,212.92)(1.062,-0.492){19}{\rule{0.936pt}{0.118pt}}
\multiput(1348.00,213.17)(21.057,-11.000){2}{\rule{0.468pt}{0.400pt}}
\multiput(1371.00,201.92)(1.746,-0.491){17}{\rule{1.460pt}{0.118pt}}
\multiput(1371.00,202.17)(30.970,-10.000){2}{\rule{0.730pt}{0.400pt}}
\end{picture}

\vspace{8mm}

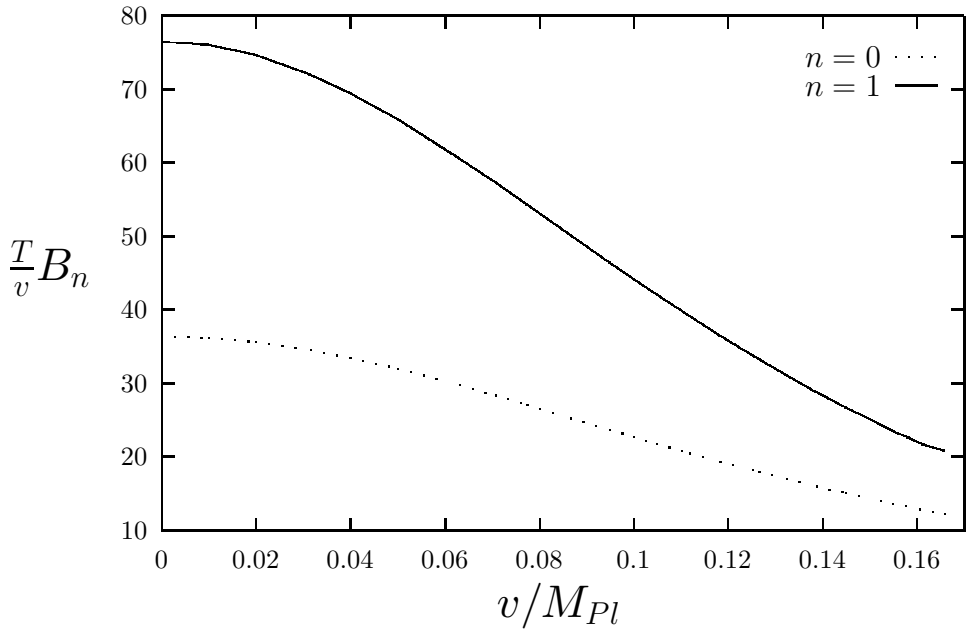
\noindent {\small Figure 5: The values of $(T/v)B_{n}$, shown versus the vacuum
expectation value of the scalar field in unit of the Planck scale, $v/M_{Pl}$.
The solid and dotted lines correspond to an $n=0$ bubble and an $n=1$ bubble,
respectively.}

\vspace{14mm}

\setlength{\unitlength}{0.240900pt}
\ifx\plotpoint\undefined\newsavebox{\plotpoint}\fi
\sbox{\plotpoint}{\rule[-0.200pt]{0.400pt}{0.400pt}}%
\begin{picture}(1500,900)(0,0)
\font\gnuplot=cmr10 at 10pt
\gnuplot
\sbox{\plotpoint}{\rule[-0.200pt]{0.400pt}{0.400pt}}%

\put(0,480){\makebox(0,0)[1]{\shortstack{\Large $\frac{T}{v}B_{n}$}}}
\put(800,-50){\makebox(0,0){\Large $\alpha$}}

\put(176.0,68.0){\rule[-0.200pt]{303.534pt}{0.400pt}}
\put(176.0,68.0){\rule[-0.200pt]{0.400pt}{194.888pt}}
\put(176.0,68.0){\rule[-0.200pt]{4.818pt}{0.400pt}}
\put(154,68){\makebox(0,0)[r]{0}}
\put(1416.0,68.0){\rule[-0.200pt]{4.818pt}{0.400pt}}
\put(176.0,169.0){\rule[-0.200pt]{4.818pt}{0.400pt}}
\put(154,169){\makebox(0,0)[r]{20}}
\put(1416.0,169.0){\rule[-0.200pt]{4.818pt}{0.400pt}}
\put(176.0,270.0){\rule[-0.200pt]{4.818pt}{0.400pt}}
\put(154,270){\makebox(0,0)[r]{40}}
\put(1416.0,270.0){\rule[-0.200pt]{4.818pt}{0.400pt}}
\put(176.0,371.0){\rule[-0.200pt]{4.818pt}{0.400pt}}
\put(154,371){\makebox(0,0)[r]{60}}
\put(1416.0,371.0){\rule[-0.200pt]{4.818pt}{0.400pt}}
\put(176.0,473.0){\rule[-0.200pt]{4.818pt}{0.400pt}}
\put(154,473){\makebox(0,0)[r]{80}}
\put(1416.0,473.0){\rule[-0.200pt]{4.818pt}{0.400pt}}
\put(176.0,574.0){\rule[-0.200pt]{4.818pt}{0.400pt}}
\put(154,574){\makebox(0,0)[r]{100}}
\put(1416.0,574.0){\rule[-0.200pt]{4.818pt}{0.400pt}}
\put(176.0,675.0){\rule[-0.200pt]{4.818pt}{0.400pt}}
\put(154,675){\makebox(0,0)[r]{120}}
\put(1416.0,675.0){\rule[-0.200pt]{4.818pt}{0.400pt}}
\put(176.0,776.0){\rule[-0.200pt]{4.818pt}{0.400pt}}
\put(154,776){\makebox(0,0)[r]{140}}
\put(1416.0,776.0){\rule[-0.200pt]{4.818pt}{0.400pt}}
\put(176.0,877.0){\rule[-0.200pt]{4.818pt}{0.400pt}}
\put(154,877){\makebox(0,0)[r]{160}}
\put(1416.0,877.0){\rule[-0.200pt]{4.818pt}{0.400pt}}
\put(176.0,68.0){\rule[-0.200pt]{0.400pt}{4.818pt}}
\put(176,23){\makebox(0,0){0}}
\put(176.0,857.0){\rule[-0.200pt]{0.400pt}{4.818pt}}
\put(386.0,68.0){\rule[-0.200pt]{0.400pt}{4.818pt}}
\put(386,23){\makebox(0,0){0.05}}
\put(386.0,857.0){\rule[-0.200pt]{0.400pt}{4.818pt}}
\put(596.0,68.0){\rule[-0.200pt]{0.400pt}{4.818pt}}
\put(596,23){\makebox(0,0){0.1}}
\put(596.0,857.0){\rule[-0.200pt]{0.400pt}{4.818pt}}
\put(806.0,68.0){\rule[-0.200pt]{0.400pt}{4.818pt}}
\put(806,23){\makebox(0,0){0.15}}
\put(806.0,857.0){\rule[-0.200pt]{0.400pt}{4.818pt}}
\put(1016.0,68.0){\rule[-0.200pt]{0.400pt}{4.818pt}}
\put(1016,23){\makebox(0,0){0.2}}
\put(1016.0,857.0){\rule[-0.200pt]{0.400pt}{4.818pt}}
\put(1226.0,68.0){\rule[-0.200pt]{0.400pt}{4.818pt}}
\put(1226,23){\makebox(0,0){0.25}}
\put(1226.0,857.0){\rule[-0.200pt]{0.400pt}{4.818pt}}
\put(1436.0,68.0){\rule[-0.200pt]{0.400pt}{4.818pt}}
\put(1436,23){\makebox(0,0){0.3}}
\put(1436.0,857.0){\rule[-0.200pt]{0.400pt}{4.818pt}}
\put(176.0,68.0){\rule[-0.200pt]{303.534pt}{0.400pt}}
\put(1436.0,68.0){\rule[-0.200pt]{0.400pt}{194.888pt}}
\put(176.0,877.0){\rule[-0.200pt]{303.534pt}{0.400pt}}
\put(176.0,68.0){\rule[-0.200pt]{0.400pt}{194.888pt}}
\put(1306,812){\makebox(0,0)[r]{$n=0$}}
\multiput(1328,812)(20.756,0.000){4}{\usebox{\plotpoint}}
\put(281,784){\usebox{\plotpoint}}
\multiput(281,784)(5.209,-20.091){5}{\usebox{\plotpoint}}
\multiput(302,703)(4.034,-20.360){5}{\usebox{\plotpoint}}
\multiput(323,597)(4.667,-20.224){4}{\usebox{\plotpoint}}
\multiput(344,506)(7.523,-19.344){3}{\usebox{\plotpoint}}
\multiput(365,452)(7.903,-19.192){3}{\usebox{\plotpoint}}
\multiput(386,401)(9.282,-18.564){2}{\usebox{\plotpoint}}
\multiput(407,359)(10.679,-17.798){2}{\usebox{\plotpoint}}
\multiput(428,324)(12.453,-16.604){2}{\usebox{\plotpoint}}
\put(460.99,281.72){\usebox{\plotpoint}}
\multiput(470,271)(15.030,-14.314){2}{\usebox{\plotpoint}}
\put(505.62,238.46){\usebox{\plotpoint}}
\multiput(512,233)(17.270,-11.513){2}{\usebox{\plotpoint}}
\multiput(554,205)(18.386,-9.631){3}{\usebox{\plotpoint}}
\multiput(596,183)(19.239,-7.787){2}{\usebox{\plotpoint}}
\multiput(638,166)(19.871,-5.993){3}{\usebox{\plotpoint}}
\multiput(701,147)(20.191,-4.807){3}{\usebox{\plotpoint}}
\multiput(764,132)(20.389,-3.884){2}{\usebox{\plotpoint}}
\multiput(806,124)(20.547,-2.935){2}{\usebox{\plotpoint}}
\multiput(848,118)(20.590,-2.615){3}{\usebox{\plotpoint}}
\multiput(911,110)(20.662,-1.968){5}{\usebox{\plotpoint}}
\multiput(1016,100)(20.722,-1.184){5}{\usebox{\plotpoint}}
\multiput(1121,94)(20.680,-1.773){5}{\usebox{\plotpoint}}
\multiput(1226,85)(20.752,-0.395){11}{\usebox{\plotpoint}}
\put(1436,81){\usebox{\plotpoint}}
\put(1306,767){\makebox(0,0)[r]{$n=1$}}
\put(1328.0,767.0){\rule[-0.200pt]{15.899pt}{0.400pt}}
\put(302,849){\usebox{\plotpoint}}
\multiput(302.58,840.60)(0.496,-2.437){39}{\rule{0.119pt}{2.024pt}}
\multiput(301.17,844.80)(21.000,-96.799){2}{\rule{0.400pt}{1.012pt}}
\multiput(323.58,742.84)(0.496,-1.443){39}{\rule{0.119pt}{1.243pt}}
\multiput(322.17,745.42)(21.000,-57.420){2}{\rule{0.400pt}{0.621pt}}
\multiput(344.58,680.71)(0.496,-2.098){39}{\rule{0.119pt}{1.757pt}}
\multiput(343.17,684.35)(21.000,-83.353){2}{\rule{0.400pt}{0.879pt}}
\multiput(365.58,596.32)(0.496,-1.298){39}{\rule{0.119pt}{1.129pt}}
\multiput(364.17,598.66)(21.000,-51.658){2}{\rule{0.400pt}{0.564pt}}
\multiput(386.58,543.03)(0.496,-1.080){39}{\rule{0.119pt}{0.957pt}}
\multiput(385.17,545.01)(21.000,-43.013){2}{\rule{0.400pt}{0.479pt}}
\multiput(407.58,498.58)(0.496,-0.910){39}{\rule{0.119pt}{0.824pt}}
\multiput(406.17,500.29)(21.000,-36.290){2}{\rule{0.400pt}{0.412pt}}
\multiput(428.58,460.98)(0.496,-0.789){39}{\rule{0.119pt}{0.729pt}}
\multiput(427.17,462.49)(21.000,-31.488){2}{\rule{0.400pt}{0.364pt}}
\multiput(449.58,428.37)(0.496,-0.668){39}{\rule{0.119pt}{0.633pt}}
\multiput(448.17,429.69)(21.000,-26.685){2}{\rule{0.400pt}{0.317pt}}
\multiput(470.58,400.61)(0.496,-0.595){39}{\rule{0.119pt}{0.576pt}}
\multiput(469.17,401.80)(21.000,-23.804){2}{\rule{0.400pt}{0.288pt}}
\multiput(491.00,376.92)(0.498,-0.496){39}{\rule{0.500pt}{0.119pt}}
\multiput(491.00,377.17)(19.962,-21.000){2}{\rule{0.250pt}{0.400pt}}
\multiput(512.00,355.92)(0.567,-0.498){71}{\rule{0.554pt}{0.120pt}}
\multiput(512.00,356.17)(40.850,-37.000){2}{\rule{0.277pt}{0.400pt}}
\multiput(554.00,318.92)(0.726,-0.497){55}{\rule{0.679pt}{0.120pt}}
\multiput(554.00,319.17)(40.590,-29.000){2}{\rule{0.340pt}{0.400pt}}
\multiput(596.00,289.92)(0.918,-0.496){43}{\rule{0.830pt}{0.120pt}}
\multiput(596.00,290.17)(40.276,-23.000){2}{\rule{0.415pt}{0.400pt}}
\multiput(638.00,266.92)(1.092,-0.497){55}{\rule{0.969pt}{0.120pt}}
\multiput(638.00,267.17)(60.989,-29.000){2}{\rule{0.484pt}{0.400pt}}
\multiput(701.00,237.92)(1.446,-0.496){41}{\rule{1.245pt}{0.120pt}}
\multiput(701.00,238.17)(60.415,-22.000){2}{\rule{0.623pt}{0.400pt}}
\multiput(764.00,215.92)(1.789,-0.492){21}{\rule{1.500pt}{0.119pt}}
\multiput(764.00,216.17)(38.887,-12.000){2}{\rule{0.750pt}{0.400pt}}
\multiput(806.00,203.92)(1.958,-0.492){19}{\rule{1.627pt}{0.118pt}}
\multiput(806.00,204.17)(38.623,-11.000){2}{\rule{0.814pt}{0.400pt}}
\multiput(848.00,192.92)(2.479,-0.493){23}{\rule{2.038pt}{0.119pt}}
\multiput(848.00,193.17)(58.769,-13.000){2}{\rule{1.019pt}{0.400pt}}
\multiput(911.00,179.92)(2.806,-0.495){35}{\rule{2.311pt}{0.119pt}}
\multiput(911.00,180.17)(100.204,-19.000){2}{\rule{1.155pt}{0.400pt}}
\multiput(1016.00,160.92)(3.839,-0.494){25}{\rule{3.100pt}{0.119pt}}
\multiput(1016.00,161.17)(98.566,-14.000){2}{\rule{1.550pt}{0.400pt}}
\multiput(1121.00,146.92)(4.930,-0.492){19}{\rule{3.918pt}{0.118pt}}
\multiput(1121.00,147.17)(96.868,-11.000){2}{\rule{1.959pt}{0.400pt}}
\multiput(1226.00,135.92)(5.944,-0.495){33}{\rule{4.767pt}{0.119pt}}
\multiput(1226.00,136.17)(200.107,-18.000){2}{\rule{2.383pt}{0.400pt}}
\end{picture}

\vspace{11mm}

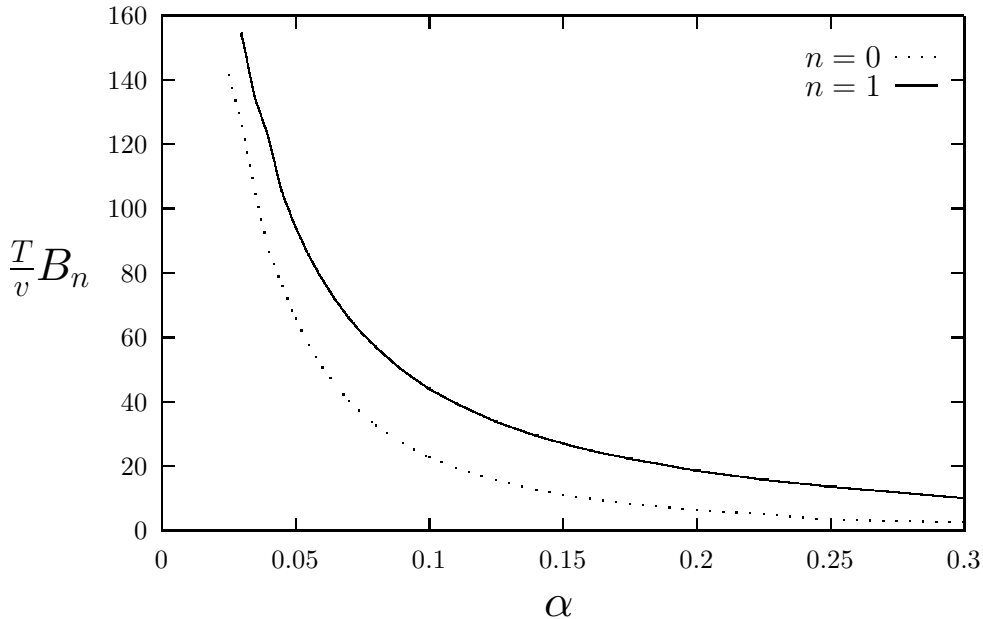
\noindent {\small Figure 6: The values of $(T/v)B_{n}$, shown versus the
parameter $\alpha$ which governs change of the shape of bubbles. The solid and 
dotted lines correspond to an $n=0$ bubble  and an $n=1$ bubble, respectively.}

\vspace{5mm}

When we have a model of a first-order phase transition with several decay 
channels, an interesting question may be which decay channel is dominant,
which is illustrated by the ratio of the two decay rates,
$\Gamma^{(1)}/\Gamma^{(0)}$. The prefactors $A_{n}$ usually are assumed to
be of the order of $m_{H}^{3}T$ at high temperature. If we neglect the
difference between $A_{0}$ and $A_{1}$, then the relative decay rate is
determined by the exponential factor, \be\label{exf}
\frac{\Gamma^{(1)}}{\Gamma^{(0)}}\sim\exp\Bigl[-\frac{v}{T}
(B^{'}_{1}-B_{0}^{'})\Bigr].
\ee
As expected, the $n=1$ bubble solutions take higher values of action, 
{\it i.e.}, $B_{1}>B_{0}$, regardless of the shape of the scalar potential
(see Fig. 6) and of the strength of gravitation (see Fig. 5). From Fig. 6,
we read that $B^{'}_{1}-B^{'}_{0}$ is large in the thin-wall limit (small
$\alpha$), because the leading  contribution of action difference
$(v/T)(B^{'}_{1}-B^{'}_{0})$ can be understood as the energy to support the
global monopole in an $n=1$ bubble, of which the long-range tail of the
global monopole, $T^{t}_{\;t}\sim v^{2}/r^{2}$, consumes the energy
proportional to the radius of the $n=1$ bubble. On the other hand, if the
bubble wall is relatively thick,  $B^{'}_{1}-B^{'}_{0}$ becomes small, and
then the nucleation of $n=1$ bubbles is not negligible. When the relative
ratio of the exponentials (\ref{exf}) is not far from the order of unity,
we should take into account the ratio of prefactors $A_{1}/A_{0}$ in order
to determine the dominant decay channel. It is difficult to compute $A_{n}$
even for an $n=0$ bubble in a flat spacetime. In a curved spacetime it has
units of energy to the fourth power and is expected to be of the order of 
$m_{H}^{3} T$. To get some information on the ratio of the prefactors, let us
consider bubbles in a flat spacetime \cite{Lin,Kim} \be
\frac{\Gamma^{(1)}}{\Gamma^{(0)}}\sim\biggl(\frac{B^{'}_{1}}{B_{0}^{'}}
\biggr)^{\frac{6}{2}}\exp\Bigl[-\frac{v}{T}(B^{'}_{1}-B_{0}^{'})\Bigr],
\ee
where the system for the fluctuations around each classical solution includes six
zero modes (three from spatial translations and another three from spatial
rotations). As explained before, $B_{1}/B_{0}-1$ represents the ratio of energy
to make a global monopole to that to generate a bubble, thus $B_{1}/B_{0}$
tends to one in the thin-wall limit and a few in the thick-wall limit. If we just
replace the values of the action in a flat spacetime to those in a curved
spacetime, we obtain several values of $\Gamma^{(1)}/\Gamma^{(0)}$, as displayed
in Table 1.

\vspace{8mm}

\begin{center}{
\begin{tabular}{|c||c|c|} \hline
$\alpha\;\backslash\;\frac{v}{T}$&1.0&2.0 \\ \hline\hline
0.3&3.32$\times10^{-2}$&1.98$\times10^{-5}$\\ \hline
0.1&3.43$\times10^{-9}$&1.59$\times10^{-18}$\\ \hline
0.03&5.23$\times10^{-13}$&1.47$\times10^{-25}$\\ \hline
\end{tabular}}
\end{center}

\vspace{1mm}

\begin{center}
{\small {Table 1. Values of $\Gamma^{(1)}/\Gamma^{(0)}$ for $\lambda=1$ and 
$v/M_{Pl}=0.1$.}}
\end{center}

\noindent
We find that monopole-bubbles become more likely to be nucleated at high 
temperature and in the relatively thick-wall case. (Remember that larger
$\alpha$ corresponds to a smaller potential barrier, which creates a bubble
with a thicker wall.) Although $B_{1}$ is always larger than $B_{0}$, there
may exist some parameter region of the scalar potential where $n=1$ bubbles
cannot be neglected.

\setcounter{section}{3}
\setcounter{equation}{0}
\begin{center}\section*{\large\bf III. Evolution of Monopole-Bubbles}
\end{center}

\indent\indent
When we consider a first-order phase transition in the framework of a
finite-temperature field theory with imaginary time, high-temperature bubbles
are given by static solutions of the Euclidean equations, and they are also
static solutions of the Lorentzian equations. It is obvious that the bubbles
start to evolve immediately after their nucleation, so a way of description is
to borrow the physics of combustion processes when the environment keeps the
temperature high enough \cite{Ste}, which is indeed the case in a flat
spacetime. Once the gravity is taken into account in the early universe, the
background universe is expanding and then it is rapidly cooled down to zero
temperature. The motion of bubbles eventually follows the zero-temperature
classical dynamics. An accurate bubble dynamics may be as follows. When the
bubble larger than the critical size is nucleated, the detonation process
induces the growth of it and simultaneously the region outside the  bubble
begins to expand due to the gravitational effect. In the next step the variation
of temperature requires the inclusion of whole  complex ingredients into the
evolution procedure of bubbles; for instance, the temperature dependence of the
the combustion processes, the effect of gravity on the classical evolution of
bubbles, the change of the effective potential due to temperature and quantum
corrections, possible reheating, the generation of Goldstone bosons and so on.
However, we already know  what actually happens to $n=0$ bubbles: if the
expansion rate of  the universe is large enough, the motion of an $n=0$ bubble
turns out to be that of a bubble governed by the classical equations of motion
at zero temperature \cite{CC,BKT,GW}. Here we suppose that the above
simplification  for $n=0$ bubbles is applied to the case of $n=1$ bubbles in a
similar manner and concentrate our interest only on the evolution of $n=1$
bubbles due to classical effects.

Our task is now reduced to solving the Einstein equations and the scalar field
equation by use of numerical analysis. In order to examine the evolution of
$n=1$ bubbles after the temperature decreases to zero, we should prepare the
initial configurations. However, they should be different from our static $n=1$
bubble solutions of the Euclidean equations because they are also static
solutions of equations of the motion after a Wick rotation. The effect due to
the change of temperature must be reflected when we prepare the initial
conditions. Suppose that the various effects mentioned above give rise to the
evolution of bubbles and then the structure of the spacetime manifold undergoes
changes, there are too many directions of perturbations to include such effects
into the initial conditions. Even under this complicated situation, we may have
several disciplines which make the problem consistent and tractable: 1. We keep
the spherical symmetry; 2. The initial configuration for the scalar fields and
the corresponding gravitational fields keeps more or less the
characteristics of those for $n=1$ static bubble solutions. Furthermore,
previous work tells us that the initial size of an O(4)-symmetric bubble at
zero temperature is larger than that of an O(3)-symmetric static bubble in the
high temperature limit, and that the O(4)-symmetric bubble expands after a
Wick rotation. (If a bubble is smaller than the static one, the amount of
surface   energy required to grow is larger than the released bulk energy
so that it is energetically favorable to shrink.)  Thus we
assume two initial configurations of the scalar amplitude $\phi$. One is a
scaled $n=1$ bubble solution in which the size of the global monopole is
also increased: \be\label{inp1} \phi(0,r)=\phi_{n=1}\Bigl({r\over
c_{1}}\Bigr)\hspace{2cm}(c_{1}>1), \ee
and the other involves the initial shrinking of monopole radius but
the expansion of the outer bubble wall: 
\be\label{inp2}
\phi(0,r)=\phi_{n=1}\left({r\over[1+(c_{2}-1)\tanh  (r/r_{{\rm
turn}}-1)]}\right) \hspace{2cm}(c_{2}>1).  
\ee

Here we solve the equations of motion by numerical calculation. As a
coordinate system in a Lorentzian spacetime, we adopt the following form,
\be\label{metric}
ds^2=-dt^2+A^2(t,\chi)d\chi^2+B^2(t,\chi)\chi^2(d\theta^2+\sin^2\theta
d\varphi^2).
\ee
With the metric (\ref{metric}), we write down the equations of motion
(\ref{eqsc}) and (\ref{ein}) as
\be\label{admp}
\ddot\phi-K\dot\phi-{\phi''\over A^2}-\Bigl(-{A'\over A}+{2B'\over
B}+{2\over\chi}\Bigr){\phi'\over A^2}+{2\delta_{n1}\phi\over \chi^2
B^2}+{dV\over d\phi}=0,
\ee
\bea\label{admhc}
-G^{t}_{\;t}&\equiv&
(2K-3K^{\theta}_{\;\theta})K^{\theta}_{\;\theta}
-\frac{2}{A^{2}}\frac{B^{''}}{B}
+\frac{B'}{A^{2}B^2}\biggl(2\frac{A^{'}}{A}-\frac{B^{'}}{B}\biggr)
+\frac{2}{\chi A^{2}}\biggl(\frac{A^{'}}{A}-3\frac{B^{'}}{B}\biggr)
\nonumber\\
& &-\frac{1}{\chi^{2}}\biggl(\frac{1}{A^{2}}-\frac{1}{B^{2}}\biggr)
=8\pi G\Bigl({\dot\phi^2\over2}+{\phi'^2\over2A^2}+{\delta_{n1}\phi^2\over
\chi^{2}B^2}+V\Bigr)
\eea
\be\label{admmc}
\frac{1}{2}G^{t}_{\;\chi}\equiv
K^{\theta '}_{\;\theta}+\biggl(\frac{B^{'}}{B}+\frac{1}{\chi}\biggr)
(3K^{\theta}_{\;\theta}-K)
=4\pi G\dot\phi\phi'
\ee
\be\label{admk}
\frac{1}{2}(G^{\chi}_{\;\chi}+G^{\theta}_{\;\theta}
+G^{\varphi}_{\;\varphi}-G^t_{\; t})\equiv 
\dot{K}-K^{2}+4KK^{\theta}_{\;\theta}-6K^{\theta^{2}}_{\;\theta}
=8\pi G(\dot\phi^2-V),
\ee
where the overdot $\dot{~}$ and the prime $'$ stand for the partial 
derivative with respect to $t$ and $\chi$ in Eq.(\ref{metric}),
respectively. Following Ref.\cite{ADM}, we have introduced the extrinsic
curvature tensor of a $t=$ const  hypersurface, whose components are given
by \be\label{kk}
K^{\chi}_{\;\chi}=-{\dot A\over A},\hspace{10mm}K^{\theta}_{\;\theta}
=K^{\varphi}_{\;\varphi}=-{\dot B\over B},
\ee
and we have denoted its trace by $K\equiv K^i_{~i}$.

For initial data for the metric, we assume an asymptotically flat de Sitter
metric just for a technical reason. Although the metric of a Euclidean 
spacetime is asymptotically closed de Sitter spacetime, the effect of the
spatial curvature is not so important as long as a bubble is smaller than
the horizon, and hence our treatment can be verified. Specifically, we
suppose $A(t=0,r)=B(t=0,r)=1$ and solve the constraint equations
(\ref{admhc}) and (\ref{admmc}) to determine $K(t=0,r)$ and
$K^{\theta}_{~\theta}(t=0,r)$. This treatment is suitable for this system
because we obtain
\begin{equation}
-{K\over3}\approx-K^{\theta}_{\;\theta}\approx\sqrt{{8\pi G\over3}
(-T^t_{\;t})},
\end{equation}
which approaches zero as $r$ increases; we can construct an
asymptotically flat spacetime without iterative integration. We have also 
assumed $K(t=0,r)<0$, which means that every point in the spacetime is
locally expanding. As for the configuration of the scalar field, we have
supposed Eq.(\ref{inp1}) or (\ref{inp2}) and $\dot\Phi(t=0,r)=0$.

In order to solve the dynamical equations, we use a finite difference 
method with 1000 meshes. The evolution of a bubble is depicted by 5
dynamical variables, $A,\; B,\; K,\; K^{\theta}_{\;\theta}$ and $\phi$.
Equations (\ref{kk}), (\ref{admk}) and (\ref {admp}) provide the next
time-step of $A,~B,~K$ and $\phi$, respectively. At each step, we integrate
(\ref{admmc}) in the radial direction to obtain $K^{\theta}_{\;\theta}$. In
this way we have reduced spatial derivatives  appearing in the equations,
which may become seeds for numerical instability. The Hamiltonian
constraint equation (\ref{admhc}) remains unsolved during the evolution and
is used for checking numerical accuracy. We keep numerical error less than
1 percent throughout calculations.

The results of our numerical computations are summarized as follows. Figure 7
shows that the outer wall of an $n=1$ bubble starts to expand, whether the
bubble wall is thin or thick. Figure 7 also tells us that the velocity of the
outer wall increases as time elapses. As the bubble grows, the wall becomes
thinner and the energy accumulated inside the bubble moves out to support the
expansion of the outer bubble wall (see Fig. 7-(b)). Therefore, after time
elapses sufficiently, the motion of the outer wall can be modeled by that of the
extremely thin wall even for any bubble with initially a thick wall. We trace
the time-evolution of the position of $\phi=0.5v$ for the outer wall, as shown
in Fig. 8. Its trajectory looks like a hyperbola, similar to that of the $n=0$
bubble wall.

\vspace{3mm}

\setlength{\unitlength}{0.240900pt}
\ifx\plotpoint\undefined\newsavebox{\plotpoint}\fi
\sbox{\plotpoint}{\rule[-0.200pt]{0.400pt}{0.400pt}}%


\vspace{11mm}

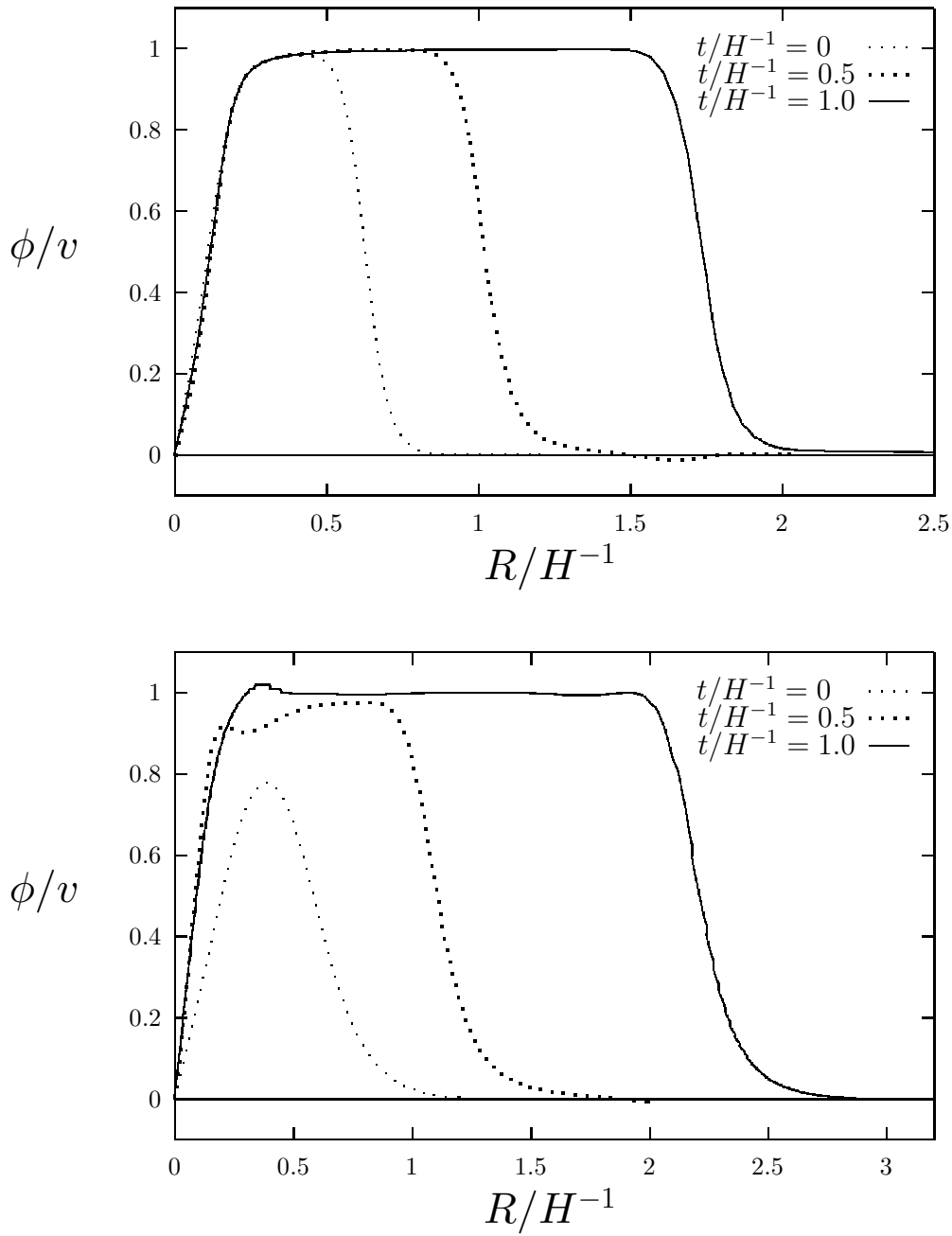
\noindent {\small Figure 7: $n=1$ bubble profiles for fixed $\lambda=1$ and
$v/M_{Pl}=0.1$ associated with the classical evolution ($c_{1}=1.2$).
$t/H^{-1}=0$, 0.5, 1.0 are shown as dotted, dashed and solid lines,
respectively. (a) thin-wall bubble of $\alpha=0.1$, (b) thick-wall bubble
of $\alpha=0.3$.}

\vspace{3mm}

\setlength{\unitlength}{0.240900pt}
\ifx\plotpoint\undefined\newsavebox{\plotpoint}\fi
\sbox{\plotpoint}{\rule[-0.200pt]{0.400pt}{0.400pt}}%
%

\vspace{11mm}

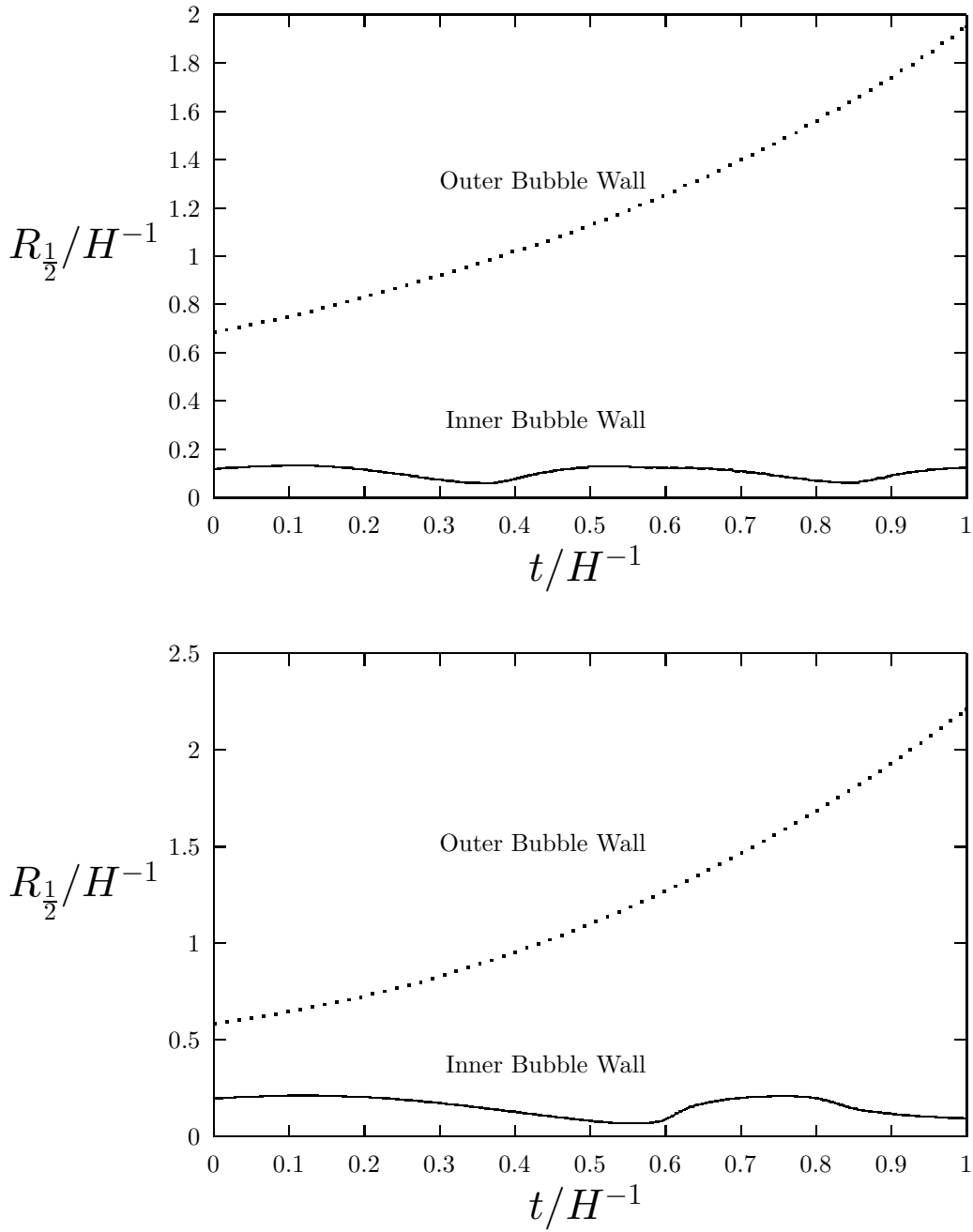
\noindent {\small Figure 8: Time evolution of the $n=1$ bubble radii for
$\lambda=1$, $v/M_{Pl}=0.1$ and $c_{1}=1.2$. (a) thin-wall bubble of
$\alpha=0.1$,  (b) thick-wall bubble of $\alpha=0.3$.}

\vspace{5mm}

An interesting physical quantity at the moment is the terminal velocity in
terms of the outer expanding coordinates. For an $n=0$ bubble it is computed in
the thin-wall approximation and, because of the gravitational effect, the
terminal velocity of the wall is found to be smaller than the light velocity
\cite{BKT}. For an $n=1$ bubble, because the spacetime between the inner wall
and the outer wall  is described by Eq.(\ref{mout}) and the spacetime outside
the outer wall is de Sitter spacetime (\ref{dS}), we obtain a junction condition,
\be\label{junc}
\epsilon^+\sqrt{\Bigl(\frac{dR}{d\tau}\Bigr)^{2}+1-\frac{8}{3}\pi
GV(0)R^{2}}- \epsilon^-\sqrt{\Bigl(\frac{dR}{d\tau}\Bigr)^{2}+1-8\pi
Gv^{2} +\frac{2G|M_{{\rm turn}}|}{R}}=-4\pi G\sigma R, \ee
where $R$, $\tau$, and $\sigma$ are  the circumference radius, the proper
time, and the surface energy density of the shell, respectively.
$\epsilon^+$ and $\epsilon^-$ are the signs of the angular component of
the extrinsic curvature of the 2+1-world-hypersurface. In general
$\epsilon^+$ and $\epsilon^-$ take $+1$ or $-1$, but in the present case
they are always positive. Neglecting the term of $2GM_{{\rm turn}}/r$ in
Eq.(\ref{junc}), we obtain the initial radius of the $n=1$ bubble,
\be\label{R0}
R_{n=1}(0)=\frac{R_{n=0}(0)}{2}\left(\sqrt{1-8\pi Gv^{2}}+
                   \sqrt{1-8\pi Gv^{2}+\frac{4v^2}{R_{n=0}(0)\sigma}}\right),
\ee
where $R_{n=0}(0)\equiv3\sigma/(V(0)+6\pi G\sigma^2)$ is the initial radius of the
$n=0$ bubble. When the size of a bubble is sufficiently large, Eq. (\ref{R0}) reduces to
\be
{R_{n=1}(0)\over R_{n=0}(0)}\approx \sqrt{1-8\pi Gv^{2}},
\ee
which shows why the size of $R_{n=1}(0)$ is smaller than $R_{n=0}(0)$ (Fig.
2-(a)). We numerically solve Eq.(\ref{junc}) and obtain an approximate formula
for the terminal velocity:
\be\label{terv}
{\rm v_{terminal}}\approx 1-\frac{4\pi GV(0)}{3[R_{n=0}(0)]^2},
\ee
which agrees with that of an $n=0$ bubble. Note that we did not consider the
temperature effect here in Eq.(\ref{terv}), and the inclusion of the temperature
effect can decrease the terminal velocity further \cite{Ste}. Now, let us
explain why the long-range term does not change the terminal velocity ${\rm
v_{terminal}}$. When the velocity of the outer wall reaches its terminal
velocity, an $n=1$ bubble can be approximated as a thin-wall bubble and the
energy density $T^{t}_{\;t}$ contributed from the expansion of the long-range
tail of the global monopole is
\bea\label{lrt}
T^{t}_{\;t}&=&\frac{1}{2}\dot{\phi}^{2}+\frac{1}{2A^{2}}\phi^{'2}
+\frac{1}{\chi^{2}B^{2}}\phi^{2}+V\nonumber\\
&\approx&\frac{v^{2}}{\chi^{2}B^{2}},
\eea
since the scalar field $\phi$ stays at the vacuum $v$. This contribution is
considerable at the initial stage of the evolution; however, it subsides to
zero as the $n=1$ bubble grows ($B\chi$ increases). Therefore, the energy
difference per unit surface area of the bubble becomes the same as that of
the $n=0$ bubble. We thus conclude that the terminal velocity of an $n=1$
bubble is equal to that of an $n=0$ bubble.

Another important motion for the $n=1$ bubble is of course that of the inner
wall. Figure 8 shows the trajectory of the position of $\phi=0.5v$ for the inner
wall. We find that the inner wall just oscillates, and we can expect that this
oscillation will be damped gradually. Since we assumed initial configurations
with both enlarged (Eq.(\ref{inp1})) and shrunken (Eq.(\ref{inp2}))  cores of
the global monopole, the above result implies that the global monopole inside an
$n=1$ bubble is stable against the perturbations of the scalar amplitude,
\be\label{sap}
\phi^{a}=\hat{\phi}^{a}\phi(\chi)\rightarrow\hat{\phi}^{a}(\phi(t,\chi )
+\delta\phi(t,\chi )). \ee

>From the above discussion, we summarize the evolution of bubbles as follows.
When the spherical symmetry is assumed for the scalar field, the motion of
both $n=0$ and $n=1$ bubbles is represented by the expansion of the bubble
wall. The global monopole formed in the $n=1$ bubble remains to be stable
and its long-range energy tail in Eq.(\ref{lrt}) keeps growing before
bubble percolation by consuming a part of the false vacuum energy (proportional
to the increment of bubble radius) obtained from the growth of the true vacuum
bubble (proportional to the increment of spatial volume). It explains why one
need not worry about the huge amount of energy necessary to maintain a large-size
global monopole if it is formed through the first-order phase transition. As a
monopole-bubble is larger, the ratio of the energy for keeping the global
monopole to the energy obtained through the growth of the bubble is smaller.
Furthermore, this ratio finally becomes negligible for an extremely large
bubble. We have also shown that the global monopole is a stable object during
the evolution of a spherically symmetric $n=1$ bubble; however, the stability
and its physical implication due to the distortion of the bubble or the
collision of two bubbles remain topics for future work.

One way to understand the stability of the inner bubble wall against the 
perturbation of the scalar amplitude in Eq.(\ref{sap}) is to count the number of
negative modes by considering the small fluctuation around a given $n=1$
bubble solution. Since it is too difficult to  calculate the negative modes
with the inclusion of gravity even for a $n=0$ bubble, let us attempt to do
it in a flat spacetime. Under small static fluctuations around a
spherically-symmetric bubble solution
$\delta\phi^{a}_{n}=\hat{\phi}^{a}_{n}\sum c_{k}\psi_{k}(x^{i})$, we obtain
a Schr\"{o}dinger-type equation by varying the scalar equation
(\ref{eqsc}):  \be\label{Sceq}
\biggl(-\nabla^{2}+\frac{d^{2}V}{d\phi^{2}}\biggr|_{\phi_{n}(\chi)}
\biggr)\hat{\phi}^{a}_{n}\psi_{k}(x^{i})=\lambda_{k}
\hat{\phi}^{a}_{n}\psi_{k}(x^{i}),
\ee
where $x^{i}$ ($i=1,2,3$) denote spatial Cartesian coordinates and a subscript
$n$ takes 0 or 1, corresponding to an $n=0$ or $n=1$ bubble. For an $n=0$ bubble
with $\hat{\phi}^{a}_{0}=(0,0,1)$, Eq.(\ref{Sceq}) contains a unique negative
mode of which the wave function is that of a nodeless $s$-wave \cite{CC,Col}. In
the case of $n=1$ bubbles, Eq.(\ref{Sceq}) becomes
\be
\biggl(-\nabla^{2}+\frac{2}{\chi^{2}}+\frac{d^{2}V}{d\phi^{2}}
\biggr|_{\phi_{1}(\chi)}\biggr)\psi_{k}(x^{i})=\lambda_{k}
\psi_{k}(x^{i}),
\ee
and then the lowest mode is not the nodeless $s$-wave mode but the $l=1$ mode
with a single node at $\chi=0$. For $n=0$ bubbles, the unique negative mode was
used for the explanation of the motion of its bubble wall. We already showed
through the numerical computation of $n=1$ bubbles that the inner bubble wall is
stable but the  outer bubble wall starts to evolve as time goes. Thus we are
likely to interpret this unique negative mode obtained from radial perturbation
as the one related with the motion of the outer bubble wall, albeit we need
further study to reach a definite conclusion. However, for perturbations in all
directions, it is extremely difficult to count the number of zero modes for an
$n=1$ bubble in a curved spacetime and the $n=1$ bubble can also have the
possibility of containing more than one zero modes as happens for the $n=0$
bubble \cite{BGV}.

Finally, let us discuss the case where the scale of symmetry breakdown $v$
approaches the  Planck scale $M_{Pl}$. When the de Sitter horizon is
comparable to or smaller than the radius of a bubble, the procedure of a
first-order phase transition may not follow the scenario in Ref.\cite{CD} but
drastic change occurs. For $n=0$ bubbles, one may bring up two proposed
scenarios of phase transitions: One is the one-bubble universe formed inside a
thin-wall bubble \cite{GW} and the other is the Hawking-Moss type phase
transition \cite{HM}. If we follow the viewpoint of Ref.\cite{GW} for the
$n=1$ bubble, it is probable that a bubble with a super-horizon-sized monopole is
nucleated. Once such a configuration is formed, interesting phenomena are
expected: the evolution of global monopoles \cite{GR} or the defect inflation at
the monopole site \cite{Vil,SSTM}. If we consider the model of interest with the
$SO(3)$ gauge coupling, the nucleated false vacuum island is dressed by gauge
fields, and consequently shows defect inflation and the creation of a
Schwarzschild-like wormhole \cite{MSSK,BM,Sak}.

\setcounter{section}{4}
\setcounter{equation}{0}
\begin{center}\section*{\large\bf IV. Conclusion and Discussions}
\end{center}

\indent\indent
In this paper we have studied a first-order phase transition in an
$O(3)$-symmetric model in a curved spacetime and at high temperature. We 
found a new bubble solution which describes another possible decay channel.
Different from an ordinary bubble, it contains a matter lump at the center
of the bubble, which is nothing but a global monopole supported by the
winding between the internal group $O(3)$ space and the real space. It
manifestly shows how the continuous internal symmetry of the theory can
play an important role from the beginning of bubble nucleation.

The obtained monopole-bubble ($n=1$ bubble) has the following characteristics.
First, in addition to the outer bubble wall which  distinguishes the true vacuum
region inside the bubble from the false vacuum environment, there is another
inner bubble wall which divides the core of the global monopole and the true
vacuum region inside the bubble with the long-range energy tail of the global
monopole. Since the formation of the global monopole consumes a part of the
energy obtained by the difference between false vacuum energy and true vacuum
energy, the size of the monopole-bubble is slightly larger than that of the
ordinary bubble for thick-wall bubbles. However, strikingly enough, the opposite
is true for the monopole-bubbles with sufficiently thin walls. Second, the
long-range tail of the global monopole is terminated by the cutoff ``outer bubble
wall'', and the total energy to form the global monopole is proportional to the
radius of the monopole-bubble. Furthermore, while the outer wall expands, the
global monopole itself is a stable configuration before bubble collisions. It
suggests a new mechanism for the production of global monopoles through a
first-order phase transition despite its infinite energy. The spacetime inside
the monopole-bubble is flat with solid deficit angle, but the black hole is not
produced at the center even at the Planck scale. Extending our analysis to
general models, we expect that other topological defects or non-topological
solitons can also be created through a first-order phase transition as {\it
solitonic-bubble} solutions. We should take into account a new possibility
``soliton  production at the bubble nucleation era'' in addition to the soliton
production by the horizon or bubble collisions \cite{Kib,BGV,BKVV}.

Since the action of a monopole-bubble is larger than that of an ordinary
bubble, the production rate of monopole-bubbles is exponentially
suppressed in comparison with that of ordinary bubbles at low
temperature; however, it is enhanced considerably at high temperature. We
showed that the production rate of monopole-bubbles can be comparable to
that of ordinary bubbles for some parameter range of the scalar potential,
so the first-order phase transition at high temperature described by such
models can proceed through this new decay channel by the nucleation of
monopole-bubbles. If we look at the evolution of the  monopole-bubble by
classical dynamics after the background universe  is cooled down to zero 
temperature, we can easily notice the following: 1. The outer bubble wall
immediately starts to expand and reaches the terminal velocity smaller
than light velocity, just as the case of an ordinary bubble; 2. The inner
bubble wall remains stable and the long-range tail of the global monopole
grows as the outer wall expands.

It is well known that topological defects are produced by the Kibble mechanism
or bubble collisions in the early universe \cite{Kib,BKVV}, this monopole-bubble
nucleation can be a new mechanism of producing global monopoles through a
first-order phase transition in the early universe. If we introduce the gauge
coupling, the character of a monopole inside the bubble changes from the global
one to the 't Hooft-Polyakov monopole \cite{KMS2}. It may be interesting to apply
this mechanism to the production of local monopoles and compare with the results
obtained in other ways \cite{BGV}. If the production rate of monopole-bubbles is
too large, then one must worry about monopole abundance in the early universe,
though it can be diluted by inflation. For computing the number of monopoles
which survive after the completion of the first-order phase transition, further
study of bubble collisions is needed, particularly between $n=1$ and $n=0$
bubbles, or between $n=1$ and $n=1$ bubbles. Finally, we should emphasize again
that the above procedure to nucleate new bubbles involving a soliton in its
center is due to continuous symmetry and can be generalized for any  continuous
global or local symmetry which has an appropriate winding between internal group
space and spacetime. However, it is unclear for the  discrete symmetry case in
(3+1) dimensions \cite{BP} and needs further study.

\begin{center}\section*{\large\bf Acknowledgments}\end{center}
\indent\indent
The authors are very grateful to G. 't Hooft, V.P. Nair, K. Nakao, M. Peskin,
M. Sasaki, K. Sato, T. Tanaka, E.J. Weinberg, Piljin Yi, and  Zae-young Ghim for
valuable discussions. Thanks are also due to P. Haines for correcting the
manuscript. K.M. and N.S.'s research was supported in part by the
Grant-in-aid for Scientific Research Fund of the Ministry of Education, Science
and Culture (No.06302021, No.06640412, and No.07740226), and by the Waseda
University Grant for Special Research Projects. Y.K.'s research was supported in
part by JSPS (No.93033), the KOSEF (95-0702-04-01-3, Brain Pool Program) and
the Korean Ministry of Education (BSRI-95-2413). N.S. thanks Yukawa Institute of
Theoretical Physics for financial support during his visit. Y.K. thanks Physics
Department of Hanyang University (BSRI-95-2441) and Center for Theoretical Physics of
Seoul National University for their hospitality and financial support during his 
visit.

\def\hebibliography#1{\begin{center}\subsection*{References}
\end{center}\list
  {[\arabic{enumi}]}{\settowidth\labelwidth{[#1]}
\leftmargin\labelwidth	  \advance\leftmargin\labelsep
    \usecounter{enumi}}
    \def\newblock{\hskip .11em plus .33em minus .07em}
    \sloppy\clubpenalty4000\widowpenalty4000
    \sfcode`\.=1000\relax}

\let\endhebibliography=\endlist

\begin{hebibliography}{100}
\bibitem{KL} D. A. Kirzhnits and A. D. Linde, Phys. Lett. B 42, 471
(1972); L. Dolan and R. Jackiw, Phys. Rev. D 9, 3320 (1974); S. Weinberg,
{\it ibid} 9, 3357 (1974).
\bibitem{CC} S. Coleman, Phys. Rev. D 15, 2929 (1977); C. Callan  and S.
Coleman, {\it ibid} 16, 1762 (1977).
\bibitem{CD} S. Coleman and F. De Luccia, Phys. Rev. D 21, 3305 (1980).
\bibitem{Aff} I. Affleck, Phys. Rev. Lett. 46, 306 (1981).
\bibitem{Lin} A. D. Linde, Phys. Lett. B 70, 306 (1977); {\it ibid}
100, 37 (1981); Nucl. Phys. B 216, 421 (1983).
\bibitem{Gut} A. H. Guth, Phys. Rev. D 23, 347 (1981); K. Sato, Mon. Not.
R. Astron. Soc. 195, 467 (1981); A. D. Linde, Phys. Rev. B 108, 389
(1982); A. Albrecht and P. Steinhardt, Phys. Rev. Lett. 48, 1220 (1982).
\bibitem{MSSK} K. Sato, M. Sasaki, H. Kodama and K. Maeda, Prog. Theor.
Phys. 65, 1143 (1981); K. Maeda, K. Sato, M. Sasaki and H. Kodama, 
Phys. Lett. B 108, 98 (1982).
\bibitem{Vil} A. Vilenkin, Phys. Rev. Lett. 72, 3137 (1994); 
A. D. Linde, Phys. Lett. B 327, 208 (1994);
A. D. Linde and D. Linde, Phys. Rev. D 50, 2456 (1994).
\bibitem{SSTM} N. Sakai, H. Shinkai, T. Tachizawa, and K. Maeda, Phys. Rev. D
53, 655 (1996).
\bibitem{Sak} N. Sakai, Waseda university preprint WU-AP/52/95, gr-qc/9512045.
\bibitem{Vil2} For a review, see A. Vilenkin, Phys. Rep. 121, 263 (1985)
and A. Vilenkin and E. P. S. Shellard, {\it Cosmic Strings and Other
Topological Defects}, (Cambridge, Cambridge University Press, 1994).
\bibitem{Col} S. Coleman, Nucl. Phys. B 298, 178 (1988).
\bibitem{Kim} Y. Kim, Nagoya university preprint DPNU-94-39, hep-th/9410076.
\bibitem{BV} M. Barriola and A. Vilenkin, Phys. Rev. Lett. 63 
341 (1989); D. Harari and C. Loust\'{o}, Phys. Rev. D 42, 2626 (1990).
\bibitem{Ste} P. J. Steinhardt, Phys. Rev. D 25, 2074 (1982).
\bibitem{DH} G. H. Derrick, J. Math. Phys. 5, 1252 (1964);
R. Hobart, Proc. Phys. Soc. 82, 201 (1963).
\bibitem{Gre} R. Gregory, Phys. Lett. B 215, 663 (1988); A. G. Cohen
and D. B. Kaplan, {\it ibid} 215, 67 (1988); D. Harrari  and P.
Sikivie, Phys. Rev. D 37, 3438 (1988).
\bibitem{HP} D. Harrari and A. Polychronakos, Phys. Lett. B 240, 55
(1990); G. W. Gibbons, M. E. Ortiz and F. Ruiz Ruiz, {\it ibid} B 240,
50 (1990).
\bibitem{KKK} For the rigorous proof of existence of bubble solutions in flat
spacetime, see Ref.\cite{CC} for an $n=0$ bubble, and Ref.\cite{Kim} with the
help of C. Kim, S. Kim and Y. Kim, Phys. Rev. D 47 (1993)  5434 for an
$n=1$ bubble.
\bibitem{Cosm} J. R. Gott, Astrophys. J. 288, 422 (1985);
B. Linet, Class. Quantum Grav. 7, L75 (1990); M. E. Ortiz, Phys. Rev.
D 43, 2521 (1991); C. Kim and Y. Kim, {\it ibid} 50, 1040 (1994).
\bibitem{Hog} C. J. Hogan, Phys. Lett. B 133, 172 (1983); M. S. 
Turner, E. J. Weinberg and L. M. Widrow, Phys. Rev. D 46, 2384 (1992).
\bibitem{BKT} V. A. Berezin, V. A. Kuzmin and I. I. Tkachev, Phys. Lett. B
120, 91 (1983), Phys. Rev. D 36, 2919 (1987); S. K. Blau, E.
I. Guendelman and A. H. Guth, {\it ibid} 35, 1747 (1987);
N. Sakai and K. Maeda, {\it ibid} 50, 5425 (1994).
\bibitem{GW} A. H. Guth and E. J. Weinberg, Nucl. Phys. B 212, 321
(1983).
\bibitem{ADM} R. Arnowitt, S. Deser and C. W. Misner, in {\it Gravitation:
An Introduction to Current Research} edited by L. Witten, (Wiley, New York,
1962).
\bibitem{BGV} R. Basu, A. H. Guth and A. Vilenkin, Phys. Rev. D 44,
340 (1991); J. Garriga and A. Vilenkin, {\it ibid} 47, 3265  (1993).
\bibitem{GR} E. I. Guendelman and A. Rabinowitz, Phys. Rev. D 44,
3152 (1991).
\bibitem{HM} S. W. Hawking and I. G. Moss, Phys. Lett. B 110, 35
(1982); L. G. Jensen and P. J. Steinhardt, Nucl. Phys. B 237, 176
(1984), {\it ibid} B 317, 693 (1989); D. A. Samuel and  W. A.
Hiscock, Phys. Rev. D 44, 3052 (1991).
\bibitem{BM} 
K. Lee, V. P. Nair and E. J. Weinberg, Phys. Rev. Lett. 
68, 1100 (1992), Phys. Rev. D 45, 2751 (1992); M. E. Ortiz, {\it
ibid} 45, R2586 (1992); P. Breitenlohner, P. Forg\'{a}cs, and D. Maison,
Nucl. Phys. B 383, 357 (1992); T. Tachizawa, K. Maeda and T. Torii, Phys.
Rev. D 51, 4054 (1995).
\bibitem{Kib} T. W. B. Kibble, J. Phys. A 9, 1387 (1976).
\bibitem{BKVV} J. Borrill, T. W. B. Kibble, T. Vachaspati and A. Vilenkin,
Phys. Rev. D 52, 1934 (1995).
\bibitem{KMS2} Y. Kim, K. Maeda and N. Sakai, in preparation. 
\bibitem{BP} I. V. Barashenkov and E. Yu. Panova, Physica D 69, 114 (1993).
\end{hebibliography}
\end{document}